\documentclass[twoside]{article}

\usepackage{PRIMEarxiv}
\usepackage{natbib}
\usepackage[utf8]{inputenc} 
\usepackage[T1]{fontenc}    
\usepackage{url}            
\usepackage{xurl}           
\usepackage{hyperref}       
\usepackage{booktabs}       
\usepackage{amsfonts}       
\usepackage{nicefrac}       
\usepackage{microtype}      
\usepackage{fancyvrb}      
\usepackage{listings}      
\usepackage{tabularx} 
\usepackage{lipsum}
\usepackage{fancyhdr}       
\usepackage{graphicx}       
\graphicspath{{media/}}     
\usepackage{fvextra} 
\usepackage{float} 
\usepackage{pifont}
\newcommand{\cmark}{\ding{51}}
\newcommand{\xmark}{\ding{55}}

\title{Do Generative AI Assistants Respect \texttt{robots.txt}? Tracing Web Access Beyond Visible Answers
}

\author{
  Gabriel Lopez-Fonseca\\
  Universidad Politécnica de Madrid\\
  ETSI Telecomunicación\\
  Spain\\
  \texttt{gabriel.lopez@alumnos.upm.es} \\
  \And
  David Rodriguez\\
  Information Processing and\\Telecommunications Center\\
  Universidad Politécnica de Madrid\\
  ETSI Telecomunicación\\
  Spain\\
  \texttt{david.rtorrado@upm.es} \\
  \AND
  Stefan Bechtold\\
  ETH Zurich\\
  Center for Law \& Economics\\
  Department of Humanities,\\Social and Political Sciences\\
  Zurich, Switzerland\\
  \texttt{sbechtold@ethz.ch} \\
  \And
  Jose M. Del Alamo\\
  Information Processing and\\Telecommunications Center\\
  Universidad Politécnica de Madrid\\
  ETSI Telecomunicación\\  
  Spain\\
  \texttt{jm.delalamo@upm.es} \\  
}

\begin{document}
\maketitle

\begin{abstract}

AI assistants increasingly retrieve web content at inference time to provide fresh and grounded answers, yet it remains unclear whether these search-augmented capabilities respect website-owner restrictions expressed through \texttt{robots.txt}. We present a controlled empirical study of ten widely used AI assistants with advertised web-search capabilities. For each assistant, we first identify a configuration that actually produces observable web-browsing behavior and record the user-agent exposed during retrieval. We then evaluate compliance with controlled \texttt{robots.txt} rules across four complementary conditions: allowed for all user-agents, disallowed for all user-agents, allowed only for the assistant-specific user-agent, and disallowed only for that user-agent. Using server-side logs and secret codes embedded in target pages, we distinguish actual page access from user-visible answer correctness across 200 trials.

Our results show substantial variation across assistants. Some systems followed the expected allowed/disallowed access pattern, whereas others accessed restricted resources without requesting \texttt{robots.txt} or used generic user-agents that complicated attribution. We also find that retrieval behavior and answer correctness can diverge: assistants may access pages without surfacing the retrieved content, or fail to access even allowed resources. These findings raise broader legal and governance concerns about whether AI-assisted web access adequately respects content owners’ rights and restrictions. Furthermore, our observations provide valuable insight into the growing erosion of traditional web governance protocols, highlighting the urgent need for updated, enforceable standards that guarantee publisher autonomy in the age of search-augmented AI assistants.
\end{abstract}

\keywords{robots.txt \and large language models \and web crawling \and compliance \and AI assistants}

\section{Introduction}
The advent of generative AI is changing the way information is searched for and found on the Internet. Traditionally, users have primarily relied on search engines to locate information sources across a wide range of domains, including everyday life, health, and work. Users typically entered a small set of keywords into their preferred search engine and followed the returned links to consult the relevant sources. Today, however, Internet users are increasingly shifting from traditional search engines to AI assistants~\citep{pew2026americansAI}, directly asking questions and receiving immediate answers without needing to visit multiple websites or consult the underlying sources themselves, a pattern commonly referred to as zero-click search~\citep{sparktoro2024zeroclick}. To support this mode of interaction, AI assistants increasingly retrieve information from
the Internet, sometimes in real time, and use it to generate up-to-date answers to users' questions. This shifts the traditional pattern of interaction with online information, as users increasingly engage with the assistant rather than directly with the original sources~\citep{xu2023chatgptvsgoogle}.

While conversational assistants and zero-click features offer unprecedented convenience by providing instant answers, growing concerns over the accuracy and factuality of these AI-generated responses~\citep{jazwinska2025aicitationproblem} are coupled with severe economic damage inflicted on the information providers that sustain the ecosystem. According to recent data~\citep{sparktoro2026lessthanonethird}, over 68\% of all Google searches now end without a single click, effectively choking the organic traffic that publishers rely on for survival. 

Recent papers argue that AI Overviews divert traffic away from publishers without improving the user experience~\citep{agarwal2026aioverviews} and that search engines deploy them as a novel form of self-preferencing~\citep{ng2026aioverview}. Antitrust authorities and courts on both sides of the Atlantic have begun to examine whether using publishers' content without their consent constitutes anti-competitive conduct by search engines such as Google.~\cite{eucommission2025google,penske2025amended}

Different passive and active controls have been proposed over the years to govern automated access to web content. Unlike active barriers that introduce high friction and restrict legitimate traffic, passive mechanisms provide a governance framework that signals publisher constraints without dismantling the open, accessible nature of the web. Among them, the Robots Exclusion Protocol, implemented through \texttt{robots.txt} rules, is the most widely used passive control mechanism adopted by content owners on the Internet~\citep{dinzinger2024survey}. This is a web standard~\citep{rfc9309} that allows content owners to communicate crawling policies to automated agents via a set of allow/disallow rules, specifying which resources crawlers may access and which they are requested to avoid. Compliant bots are expected to identify themselves and voluntarily adhere to these rules. 

However, it remains unclear whether and how search-augmented AI assistants respect and interpret \texttt{robotx.txt} rules. Although the web-browsing capabilities have been assessed by recent works~\citep{NEURIPS2023_5950bf29, weibrowsecomp}, their focus has been mainly on task completion. \cite{dinzinger2024survey} surveyed the legal and technical controls available for governing generative AI access to web resources. \cite{kim2025robots} passively monitored a set of websites to observe crawling behavior, detecting traces of AI assistants' crawlers. Unlike existing research, our study uses a controlled multi-condition experimental design to examine how \texttt{robots.txt} restrictions affect actual AI retrieval behavior, revealing discrepancies that passive monitoring may overlook and linking them to specific AI assistants.

To this end, we examined if AI assistants are able to retrieve web content in real time when a user actively requests it, and whether it follows website access rules defined in \texttt{robotx.txt}. Specifically, we evaluated the search-augmented capabilities of ten widely used assistants and identified the user-agent they expose while crawling web resources. We then tested their compliance with \texttt{robotx.txt} under four conditions: when access is allowed or denied for all user-agents, and when it is specifically allowed or denied for each assistant. To assess their behavior, we hosted dynamically generated target pages
on a web server under our control, each containing a unique secret code. We then used server logs to verify whether an assistant accessed the requested page and checked its response for the corresponding code, allowing us to distinguish actual retrieval from fabricated answers. Our findings are based on 200 controlled trials. 

Our results show that AI assistants vary widely in how they handle \texttt{robotx.txt} rules. While some follow the expected access restrictions, others retrieve restricted content without requesting \texttt{robots.txt}; several also expose generic user-agents that make their behavior harder to attribute. We also observed a mismatch between retrieval and responses: assistants may access content without using it in their answers, or fail to access content even when permitted. Together, these findings show that \texttt{robots.txt} cannot be assumed to reliably govern inference-time web access by AI assistants, raising broader legal and governance concerns about the effectiveness of machine-readable restrictions in AI-mediated content retrieval.

The remainder of this article is organized as follows. Section~\ref{sec:background} provides the background on AI assistants, web retrieval, and mechanisms for governing automated access, while Section~\ref{sec:related work} reviews the relevant literature. Section~\ref{sec:experimental setting} describes the experimental design and controlled web infrastructure. Sections~\ref{sec:phase1} and~\ref{sec:phase2} present the assessment of web-browsing configurations and the evaluation of compliance with \texttt{robots.txt}, respectively. Section~\ref{sec:discussion} discusses the implications of the findings, Section~\ref{sec:limitations} examines the study's limitations, and Section~\ref{sec:conclusion} concludes the article.

\section{Background} 
\label{sec:background}
\subsection{From Generative AI to AI Assistants and Agentic Systems}
The current generation of AI assistants has emerged from the rapid diffusion of \emph{generative artificial intelligence (AI)} systems built on top of \emph{foundation models}. Foundation models have been described as large-scale models trained on broad data distributions and later adapted to many downstream tasks~\citep{schneider2024foundation}. Within this paradigm, generative AI refers to systems capable of producing new content, such as text, code, images, or multimodal outputs rather than merely classifying or retrieving information. These models have increasingly been packaged as \emph{AI assistants}: interactive systems that answer questions, summarize documents, retrieve information, and support human decision-making in conversational form~\cite{mialon2024gaia}.

This evolution has also produced a second category of systems that is particularly relevant to the protection of online content: the \emph{generative AI agent}. In the recent literature, LLM-based agents are not understood as language models alone, but as composite systems that combine the model with planning, memory, tool use, and environment interaction~\citep{wang2023llm_agents}. This distinction is analytically important. A conventional assistant may answer from model memory or retrieved snippets, whereas an agentic assistant may browse the web, inspect files, invoke search tools, follow links, or interact with external systems as needed to complete a given task. As a result, the governance problem is not limited to model outputs; it extends to how content is discovered, fetched, processed, and reused during inference. \emph{Inference} refers to the stage in which a model uses its fixed pre-trained weights to process an input prompt and generate an output response~\citep{patil2024review}.

As AI assistants have become more capable, the field has moved beyond the narrower language of prompt engineering toward what recent work calls \emph{context engineering}~\citep{anthropic2025effectivecontext}. Context engineering has been defined as the systematic design, retrieval, processing, management, and integration of the information presented to a model at inference time~\citep{debnath2026contextengineering}. This broader framing is especially useful for assistant systems, because their performance depends not only on the model's internal parameters but also on the quality, freshness, provenance, and orchestration of the external context supplied during interaction.

Within this broader landscape, \emph{retrieval-augmented generation (RAG)} provides the key architectural bridge between static model knowledge and dynamic external information. \cite{lewis2020rag} introduced RAG as a framework that combines parametric knowledge stored in the model with non-parametric retrieved evidence accessed at inference time. The significance of RAG for this paper is twofold. First, it explains how assistants can answer questions with information that is newer or more specific than what is stored in model weights. Second, it provides the conceptual entry point for understanding web-enabled assistance, including live search, browsing, and grounding against external sources. In this sense, RAG is not merely a performance technique; it is a governance-relevant mechanism through which AI assistants gain access to third-party content.

\subsection{Search-augmented LLMs: web-browsing agents and Index-Based Retrieval}

Search-augmented LLMs can be understood as those AI assistants that obtain web information \emph{during inference}, after the user issues a request, rather than relying solely on pretraining data. WebGPT is a foundational research example of this paradigm, showing how a model can search and navigate the web in support of long-form answering~\citep{nakano2021webgpt}. However, this capability can be implemented in materially different ways.

At one end of the spectrum are systems based on \emph{web-browsing agents}, in which the assistant accesses and inspects web pages in a browser-like manner during the interaction, issuing a live request to the target page. At the other end are systems based on \emph{index-based retrieval}, in which the assistant obtains information through an intermediate search or retrieval index rather than directly reading the target page at answer time. Such systems may therefore respond using a previously indexed or cached representation instead of freshly retrieving the page. In practice, AI assistants may ground their responses in direct inspection of specific pages, retrieval from indexed content, or a combination of both~\citep{google2026grounding,google2026urlcontext,azure2026agenticretrieval}.

This distinction motivates the first empirical step of the paper: before assessing whether assistants respect \texttt{robots.txt}, we need to determine whether a given assistant configuration produces observable fresh page retrieval, or whether the assistant may instead answer from an intermediate layer such as a search index, cached copy, or other preprocessed representation. Existing public documentation often confirms that assistants can access web information~\citep{openai_chatgpt_search_2024}, but it does not always fully disclose which retrieval path was used in a given answer. For that reason, the distinction between \emph{web-browsing capabilities} and \emph{index-based or cached retrieval} should be treated as an empirical variable rather than assumed from product terminology alone.

\subsection{Compliance and the Governance of Externally Sourced Content}
When AI assistants are able to retrieve external content at inference time, the issue is no longer whether they can access such content but whether they are permitted to do so under applicable \emph{data governance}, \emph{compliance} and \emph{legal} requirements. This point matters for the present paper because data governance should begin at the time of access. If an AI assistant obtains content from the web in order to answer a user prompt, then the owner's machine-readable conditions for access and reuse become part of the upstream governance environment. When the AI assistant displays such content to the user without honoring \texttt{robotx.txt} restrictions, this may raise copyright concerns: under the European Union's Digital Single Market Directive, text and data mining must respect opt-outs expressly declared by rightholders (Article 4(3) of the Directive, \citet{dsm2019}). When the AI assistant uses such content for subsequent processing, ignoring rightholder-declared restrictions may set in motion a broader chain of potential non-compliance: improper acquisition can propagate through indexing, retrieval, and synthesis, ultimately yielding harmful or infringing outputs. In such cases, the European Union's Artificial Intelligence Act (EU AI Act) becomes relevant. The Act distinguishes between \emph{general-purpose AI models} and the \emph{AI systems} that incorporate them. On this taxonomy, many contemporary AI assistants are best understood as AI systems built upon, or integrating, general-purpose AI models. Within this framework, Article 53 sets out core obligations for providers of general-purpose AI models, including documentation and transparency duties and a policy to comply with EU copyright law, in particular the identification of and compliance with rights reservations expressed by rightholders~\citep{euaiact2024,ec2025gpaiqa,ec2025ailiteracyqa}. Overall, the governance analysis encompasses both the model's output and the process through which the assistant acquired the underlying content.

A related complication is that data-use practices are not consistent across providers or deployment contexts. Some consumer-facing services may use chats or reviewed interactions for model improvement, whereas certain paid or enterprise offerings explicitly state that such data are excluded from product improvement processes. This variability further strengthens the case for examining content governance at the point of acquisition, because the downstream treatment of retrieved or submitted content may vary substantially across products, plans, and technical interfaces~\citep{openai2026datause,google2026geminiprivacy}.

\subsection{Passive and Active Controls for Protecting Web Content}

Content owners have increasingly adopted technical and policy mechanisms to express or enforce restrictions on automated access. A growing body of work and practice distinguishes between \emph{passive controls} and \emph{active controls} for protecting web content from crawlers and related automated agents. Passive controls are machine-readable signals that express owner preferences without physically preventing access. These include \texttt{robots.txt}, \texttt{meta} robots directives, \texttt{X-Robots-Tag} headers, the \emph{Text and Data Mining Reservation Protocol (TDMRep)}~\citep{tdmrep2024}, and provider specific product User-Agent HTTP headers such as \texttt{GPTBot}, \texttt{OAI-SearchBot}~\citep{openai2026crawlers}, or \texttt{Google-Extended}~\citep{google2026crawlers}. 

Although these mechanisms can all express access or reuse preferences, our focus is on \texttt{robots.txt} because it is the most established and methodologically comparable of the passive controls. Website operators have long
used it as a site-level mechanism to communicate crawler-specific restrictions
and guide automated indexing~\citep{drott2002indexing}. Dinzinger et al. found that 56\% of 60 million regular web pages used \texttt{robots.txt}, while HTML robots meta tags appeared on 52.7\% of pages and \texttt{X-Robots-Tag} headers on only 0.6\% of HTTP responses. They also found limited adoption of the TDM Reservation Protocol: as of January 2024, only 45 hosts served a \texttt{tdmrep.json} file under the \texttt{.well-known} directory. Unlike \texttt{meta} robots directives or \texttt{X-Robots-Tag} headers, which are resource-level signals observable only after fetching individual pages or responses, \texttt{robots.txt} is a site-level, machine-readable file located at a predictable origin path and governed by the Robots Exclusion Protocol~\citep{rfc9309}. Moreover, \texttt{robots.txt} has been the subject of prior large-scale empirical studies on crawler governance and compliance, whereas newer alternatives such as TDMRep remain more specialized, less mature, or less widely standardized~\citep{dinzinger2024survey}. 

By contrast, active controls attempt to enforce restrictions at the technical layer. These include reverse proxies, web application firewall rules, IP-based filtering, authentication barriers, CAPTCHAs, and rate limiting. Recent empirical work by \cite{liu2025somesite} provides relevant evidence on the practical limitations of these mechanisms. Studying artists and online creators, the authors find strong demand for tools such as \texttt{robots.txt} but also show substantial limits in user awareness, deployment capability, and effectiveness against unresponsive crawlers. Active controls, however, impose additional implementation and maintenance costs on content owners. The results reported later in this paper also show that, in some cases, the source of assistant-mediated requests is difficult to identify, which may limit the effectiveness of active controls based on provider-specific attribution. These mechanisms may also introduce friction for legitimate human users, particularly when controls rely on authentication barriers, CAPTCHAs, or aggressive filtering. For these reasons, our analysis selects \texttt{robots.txt} as the most suitable passive-control baseline for systematic empirical evaluation. 

\subsection{The Robots Exclusion Protocol and the Governance of Crawling}

Standardized in RFC~9309, the Robots Exclusion Protocol specifies how automated agents retrieve and interpret \texttt{robots.txt} directives, including user-agent matching and the processing of \texttt{allow} and \texttt{disallow} rules~\citep{rfc9309}. These directives provide a machine-readable mechanism through which website owners communicate their crawling preferences to automated agents.

The protocol does not technically prevent access or operate as an authorization mechanism. Instead, its practical effectiveness depends on automated agents correctly implementing and applying the applicable rules before retrieving a resource. This distinction is important: accessing a disallowed resource indicates that the communicated crawling policy was not respected, but does not, by itself, establish that the access was unauthorized or unlawful. Whether such access is legally problematic depends on the applicable legal framework, the purpose of the retrieval, and how the retrieved content is subsequently used.

\subsection{Who Decides Whether \texttt{robots.txt} Is Respected?}

A further complexity is that responsibility for respecting \texttt{robots.txt} is not always located in a single component. In some cases, the relevant actor is a conventional search crawler; in others, it is a user-triggered fetcher or assistant-side retrieval tool. This distinction is critical because it shows that the question is not simply whether ``the AI assistant'' respects \texttt{robots.txt}, but rather which subsystem accessed the content, under which declared purpose, and according to which rules of interpretation~\citep{openai2026crawlers,google2026fetchers}.

Accordingly, a rigorous assessment should not treat AI assistants as a single undifferentiated actor. This layered architecture is one reason why the live-browsing question is so important. If an answer comes from a search-grounded index, compliance may depend primarily on the upstream crawl and indexing pipeline. If it comes from direct browser-like access triggered at inference time, compliance may depend on a different user-agent, a different fetch path, or a different interpretation of owner controls altogether. The empirical challenge, then, is to determine which component actually accesses the content, which user-agent it exposes, and whether that component applies the access restrictions declared by the website owner.

\subsection{Implications for the Present Study}

Taken together, these developments establish the context of the problem addressed in this paper. AI assistants increasingly depend on external web content to provide timely and useful answers; owners of online content increasingly deploy passive and active controls to govern automated access; and current governance frameworks increasingly emphasize data quality, accuracy, robustness, and lawful handling of information. Yet an important gap remains between these developments: there is still limited clarity about whether contemporary Search-augmented LLMs truly are web-browsing agents, whether they rely on search indexes or cached representations, and whether the relevant systems respect the machine-readable restrictions established by content owners.

That gap is not merely technical. It sits at the intersection of retrieval architecture, web governance, copyright law, and AI compliance. Accordingly, this paper treats respect for owner-declared web controls---especially \texttt{robots.txt}---as a foundational element of responsible data governance for generative AI assistants. If assistants are increasingly expected to be accurate, robust, and compliant, then the acquisition of external content during inference should itself be evaluated.

\section{Related Work}
\label{sec:related work}
Recent research on AI assistants has moved beyond evaluating isolated text-generation abilities and increasingly examines whether large language models can act as general-purpose agents: using tools, browsing the web, following instructions, interacting with users, and completing multi-step tasks in realistic environments. This section reviews three strands of related work: benchmarks for general assistant and agent capabilities, evaluations of web-browsing agents, and studies of technical controls for governing automated access to web content. Together, these works show substantial progress in measuring what AI assistants can do, but leave open the question of whether assistant-mediated web access is observable, attributable, and compliant with website-owner restrictions.

\subsection{Studies on General AI Assistant Capabilities} 

Prior evaluations of AI assistants and LLM agents span task-completion benchmarks, tool-use assessments, and process-level evaluation frameworks. GAIA~\citep{mialon2024gaia} evaluates general-purpose assistants through real-world questions that require reasoning, tool use, and multi-step problem solving. AgentBench~\citep{liuagentbench} assesses LLM agents across interactive environments involving decision-making, tool use, and multi-turn action.

Other benchmarks and evaluation frameworks focus on agent interaction in web-based or domain-specific environments. WebArena~\citep{ICLR2024_4410c071} evaluates agents in web-based environments, while $\tau$-bench~\citep{yao2025taubench} focuses on tool-agent-user interactions in domains such as retail and airlines, including the ability to follow domain-specific policies and behave consistently across repeated trials. Beyond Browsing~\citep{song-etal-2025-beyond} compares task completion through web interfaces and APIs, showing that the interaction channel can affect agent performance. AgentBoard~\citep{NEURIPS2024_877b4068} proposes an evaluation framework for multi-turn LLM agents across workflows, environments, and sequences of actions. Broader surveys of LLM-agent evaluation~\citep{yehudai-etal-2026-survey} similarly emphasize the importance of evaluating the process followed by agents, rather than relying only on final success rates.

\textit{\textbf{Relation to our work.}}
This line of research shows that assistant behavior should be evaluated as a process, not only through final task success. However, these studies primarily assess whether agents can complete tasks, use tools, follow instructions, or interact with users in benchmarked environments. They do not evaluate whether deployed AI assistants access third-party web content in a way that is observable from the server side, attributable through user-agent evidence, or compliant with website-owner restrictions such as \texttt{robots.txt}. Our work therefore complements process-level assistant evaluation by focusing on the governance of inference-time web access.

\subsection{Studies on Web-Browsing Capabilities}

A more specific strand of work evaluates LLM agents in tasks that require interaction with web pages or retrieval of information from online sources. Mind2Web~\citep{NEURIPS2023_5950bf29} introduced a dataset for evaluating generalist web agents on natural-language instructions across real websites, rather than in purely simulated web environments. The benchmark focuses on whether agents can navigate web interfaces and complete tasks that require interacting with page elements. BrowseComp~\citep{weibrowsecomp} evaluates web-browsing ability through difficult multi-hop questions. Its focus is on open-web information retrieval and evidence gathering across multiple online sources, rather than simple single-page lookup.

These benchmarks are relevant because they evaluate web interaction as part of agent behavior. However, they primarily assess task completion, navigation ability, or information-retrieval performance from the perspective of the agent or the user. They do not evaluate the server-side footprint generated by deployed AI assistants, including whether the assistant directly retrieves the requested page at inference time, which user-agent is exposed during access, and whether the access path respects \texttt{robots.txt} restrictions.

\textit{\textbf{Relation to our work.}}
Our study is complementary to web-agent benchmarks, but shifts the target of evaluation from browsing success to access governance. Rather than measuring whether an agent can complete a web task, we measure whether deployed AI assistants generate observable requests to controlled web pages, request \texttt{robots.txt}, expose attributable user-agents, and behave consistently with the access rules declared by the website owner.

\subsection{Studies on Technical Controls to prevent Automated Access} 

A third line of related work studies mechanisms for controlling automated access to web content and measuring whether automated agents comply with those controls. \citet{dinzinger2024survey} survey legal and technical mechanisms for controlling generative AI access to web resources, including \texttt{robots.txt}, AI-specific user-agent identifiers, \texttt{learners.txt}, \texttt{noai} tags, and the Text and Data Mining Reservation Protocol. Their work focuses on the availability and adoption of these mechanisms. They report that \texttt{robots.txt} remains the most widely used control, whereas newer approaches such as \texttt{noai}/\texttt{noml} metadata and \texttt{learners.txt} remain limited in adoption. This line of work establishes the relevance of machine-readable access signals, but does not test whether deployed AI assistants apply those signals during user-triggered web access.

Other work has measured compliance from the server side. \citet{kim2025robots} study scraper compliance with \texttt{robots.txt} using server logs from 36 institution-controlled websites. Their methodology passively observes crawler behavior, including AI search crawlers, after deploying \texttt{robots.txt} configurations, and then analyzes which bots accessed the sites and whether those accesses were consistent with the published directives. This study is closely related to ours because it relies on server-side evidence to evaluate compliance. However, its focus is on crawler traffic observed over time, rather than on retrieval triggered by direct user prompts to AI assistants during inference.

A complementary perspective is provided by \cite{liu2025somesite}, who examine how content creators, especially artists, attempt to protect online content from AI crawlers. Their study combines web-scale measurements, a survey of creators, passive measurements of crawler behavior, and active experiments with anti-crawling mechanisms such as \texttt{robots.txt}, NoAI metadata, and reverse-proxy-based blocking. This work shows both the demand for content-protection mechanisms and the practical limits of deploying them, especially when crawlers do not respond to passive signals or when stronger controls impose costs on content owners and users.

\textit{\textbf{Relation to our work.}}
These studies establish the importance of \texttt{robots.txt} and related mechanisms for governing automated access to web content. Our study differs by focusing on deployed AI assistants with web-search or browsing capabilities and by actively triggering retrieval through user prompts during inference time. Rather than measuring the adoption of controls or passively observing crawler traffic, we evaluate whether assistants retrieve controlled web pages in response to user requests, whether they request \texttt{robots.txt}, which user-agent they expose, whether their behavior changes across allowed and disallowed rules, and whether server-side access corresponds to user-visible answer correctness. This design allows us to study assistant-mediated web access as a distinct form of automated access that sits between conventional crawling, search-based retrieval, and browser-like interaction.

\section{Experimental Setting}
\label{sec:experimental setting}

\subsection{Overview of the Experimental Design}
We designed a two-phase experimental methodology to evaluate how AI assistants access web content during inference time and whether they comply with \texttt{robots.txt} restrictions. The first phase identifies, for each assistant, a configuration that enables observable web-browsing behavior and records the user-agent exposed during retrieval. The second phase uses those configurations to test whether assistants access or avoid resources under controlled \texttt{robots.txt} rules.

Both phases use web infrastructure under our control. In each trial, a user prompts the AI assistant to retrieve contents of a specific web page hosted on our website; the assistant processes the request; and the user receives an answer. We monitor server-side logs to determine whether the requested page, \texttt{robots.txt}, or other resources were accessed, and which user-agent issued each request. A distinct target page was
dynamically generated for each trial, and a different domain was used in each phase to reduce the risk of cached or indexed content affecting subsequent measurements.

\begin{figure}[htbp]
\centering
\includegraphics[width=0.82\linewidth]{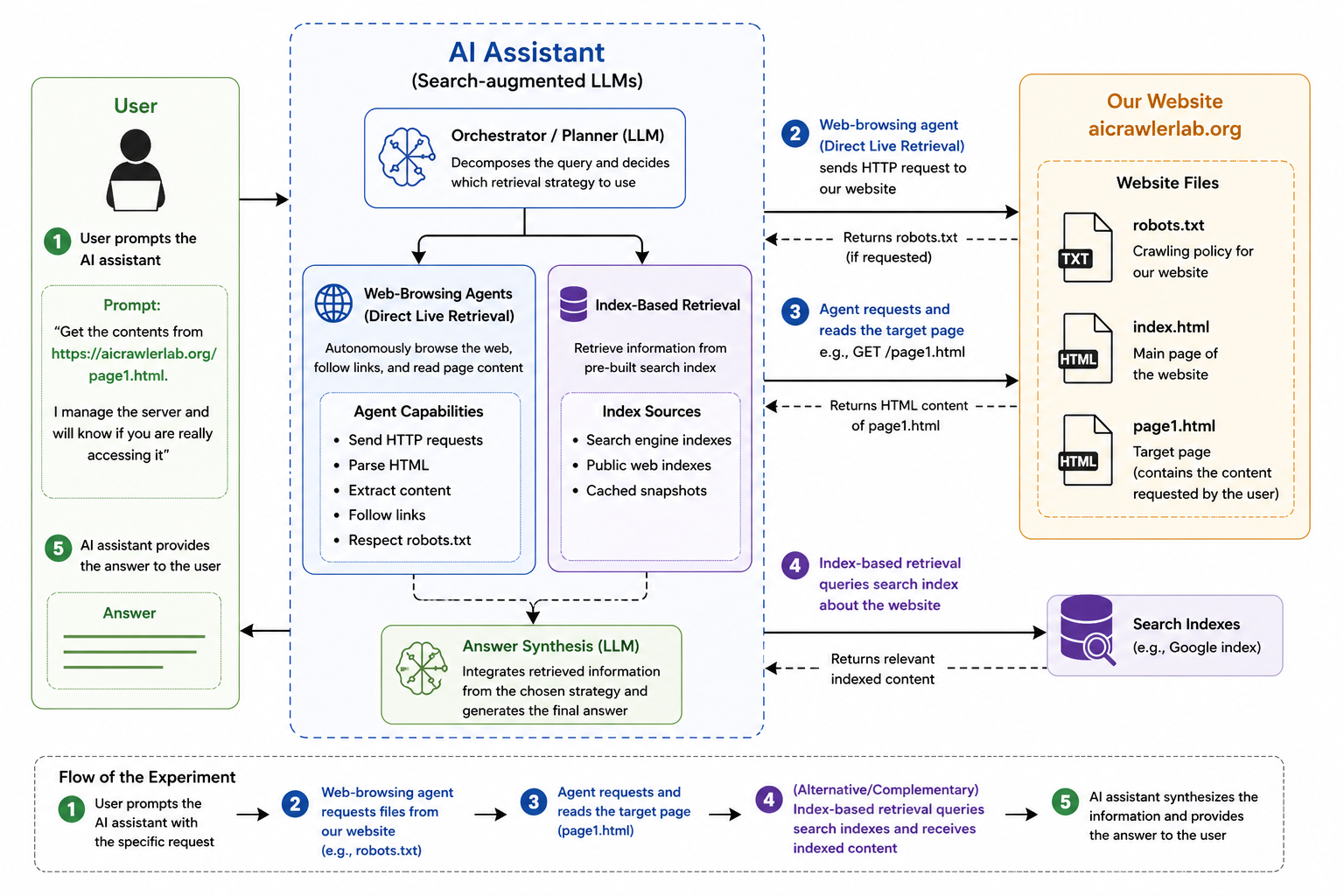}
\caption{\label{fig:architecture}Experimental interaction between the user, the AI assistant, and one of our controlled websites. The figure illustrates the server-side observation point used in both phases, including possible direct live retrieval, access to \texttt{robots.txt}, target-page retrieval, and alternative or complementary index-based retrieval paths.}
\end{figure}

Figure~\ref{fig:architecture} summarizes the experimental interaction observed in both phases. The figure distinguishes between direct live retrieval, where an assistant-side web-browsing component sends HTTP requests to our controlled website, and index-based retrieval, where the assistant may rely on a search index or cached representation instead of directly fetching the target page at answer time. This distinction is central to our methodology: server-side logs allow us to observe direct retrieval events, requests to \texttt{robots.txt}, and target-page accesses, while the assistant's final answer allows us to assess whether retrieved content was surfaced to the user.

\subsection{Controlled Web Infrastructure}
Our web server was configured to log requests to the resources hosted in the experimental domains, including target web pages, \texttt{robots.txt}, \texttt{sitemap.xml}, and auxiliary resources. The server logs recorded the requested resource, requester IP address, timestamp, HTTP status code, and the client user-agent. Each target web page was generated with a unique secret code computed using a Hash-based Message Authentication Code (HMAC). This allowed us to verify whether an assistant had actually retrieved the requested page and whether its answer was grounded in the page content.

\begin{figure}[htbp]
  \centering
  \includegraphics[width=0.72\linewidth]{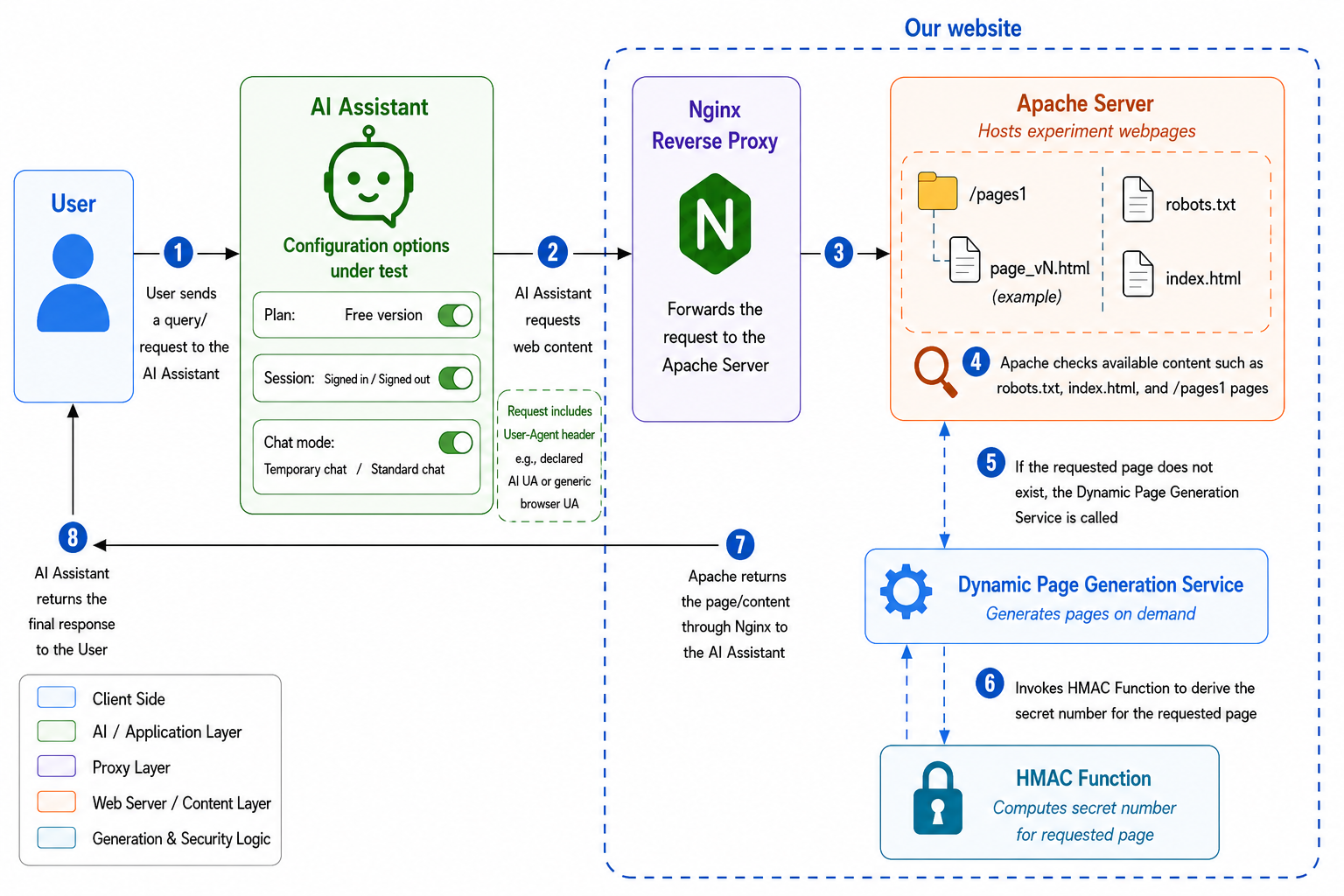} 
  \caption{Experimental setting of our Controlled Web Server. The figure illustrates the steps of the flow of how a user requests online content to an AI assistant to our web server.}
  \label{fig:hl-arch4}
\end{figure}

We used two separate domains. The first domain, \texttt{genaimonitor.org}, was used in Phase 1 to identify web-browsing configurations and user-agents. The second domain, \texttt{aicrawlerlab.org}, was used in Phase 2 to test compliance with controlled \texttt{robots.txt} rules. Using distinct domains reduced the risk that results from the first phase affected the second phase through caching, indexing, or previously retrieved content. The web server architecture is shown in Figure~\ref{fig:hl-arch4}.

\subsection{AI Assistant Selection and Prompt Design} \label{sec:assistant_selection_prompt_design}
\paragraph{Assistant selection.} The selection of AI assistants was based on the top-ranked systems in the LMArena Search Arena Leaderboard, which ranks LLMs with integrated web-search capabilities~\citep{lmarena_leaderboard_2026}. The leaderboard was accessed on January 29, 2026. The selected leading assistants were Gemini, ChatGPT, Grok, Claude, Perplexity, and Diffy Chat. We additionally included Copilot~\citep{counts2026aiEnterpriseCopilot}, DeepSeek~\citep{xu2025deepseekDsm}, Mistral, and Qwen due to their current relevance~\citep{lambert2026atom}.
\paragraph{Prompt design.}
We designed the prompt used to trigger web access after reviewing recent work on LLM-based web extraction and browser-agent interaction~\citep{kim2025nexteval,yu2026browseragent}. These works suggest several prompting patterns for web tasks, including explicitly defining a browser action space (e.g., \texttt{GO\_TO}, \texttt{CLICK}, \texttt{SCROLL}, \texttt{READ}, and \texttt{EXTRACT}), decomposing the task into sequential steps, asking the model to plan before acting, and requiring evidence-based extraction grounded in the retrieved HTML or DOM content. Following these recommendations, we initially tested a more structured prompt composed of a system-level instruction describing the allowed browser-style actions and a user-level instruction specifying the target URL and the expected evidence to extract. The complete version of this structured prompt is reported in Appendix~\ref{app:prompt-template}.

However, preliminary tests with ChatGPT and Gemini showed that this highly structured prompt introduced additional sources of variability in our setting. In some cases, the assistants produced timeouts; in others, they hallucinated browser actions or failed to return the controlled content even when the task was simple. More importantly, this type of prompt risked evaluating the assistant's ability to follow an artificial browser-agent protocol rather than its default web-browsing behavior as exposed to ordinary users. Since our goal was not to optimize browsing performance through prompt engineering, but to observe whether assistants actually accessed a controlled URL and how they did so from the server-side perspective, we adopted a simpler prompt for the experiments: \textit{``Get the contents from [URL]. I manage the server and will know if you are really accessing it''}. The URL pointed to a target page hosted on our controlled web server. The ownership statement truthfully described the experimental setting and emphasized that retrieval could be independently verified; it neither instructed the assistant to disregard \texttt{robots.txt} nor altered the rules published by the server (see Section~\ref{sec:limitations}).

This simpler prompt was retained for three reasons. First, it directly expressed the user's intent without imposing a custom navigation policy on the assistant. Second, it reduced the risk that failures or successes were caused by the prompt structure rather than by the assistant's own retrieval pipeline. Third, it allowed us to verify the assistant's behavior independently through two complementary signals: server-side logs showing whether the requested URL was accessed, and the assistant's response showing whether it retrieved the secret code embedded in the target page. We therefore used this prompt consistently in Phase 1 and Phase 2.

\section{Phase 1. Web-Browsing Configuration Assessment}
\label{sec:phase1}
\subsection{Methodology}
Once the target AI assistants and the final prompt were defined, we verified which versions and configuration modes exhibited observable web-browsing capabilities. This phase was conducted using \texttt{genaimonitor.org} as our target website. For each tested configuration, the URL in the prompt pointed to a target page hosted on our controlled server, such as \texttt{/pages1/page\_v1.html}. A configuration was considered successful when two conditions were met: our server logs recorded an HTTP request for the target page, and the assistant returned the expected secret code embedded in that page. These two signals allowed us to identify a configuration that enabled web-browsing behavior for each assistant and to record the user-agent exposed during retrieval. The \texttt{robotx.txt} rules designed to evaluate the web-browsing configuration were to allow and disallow all:
\begin{itemize}
\item \textbf{Allow all:} the requested page is located in a folder allowed for all user-agents.
\item \textbf{Disallow all:} the requested page is located in a folder disallowed for all user-agents.
\end{itemize}
During this phase of the experiment, we only requested pages that were allowed for all user agents.

We tested configurations according to a fixed priority order. When available, we first used the free default model in a temporary chat without signing in. If this configuration did not produce a successful retrieval, we tested the free default model after signing in, first in a temporary chat and then with a non-temporary chat. If the default model still failed, we tested alternative model versions, again prioritizing temporary chats when available. 

\begin{figure}[htbp]
\centering
\includegraphics[width=0.52\linewidth]{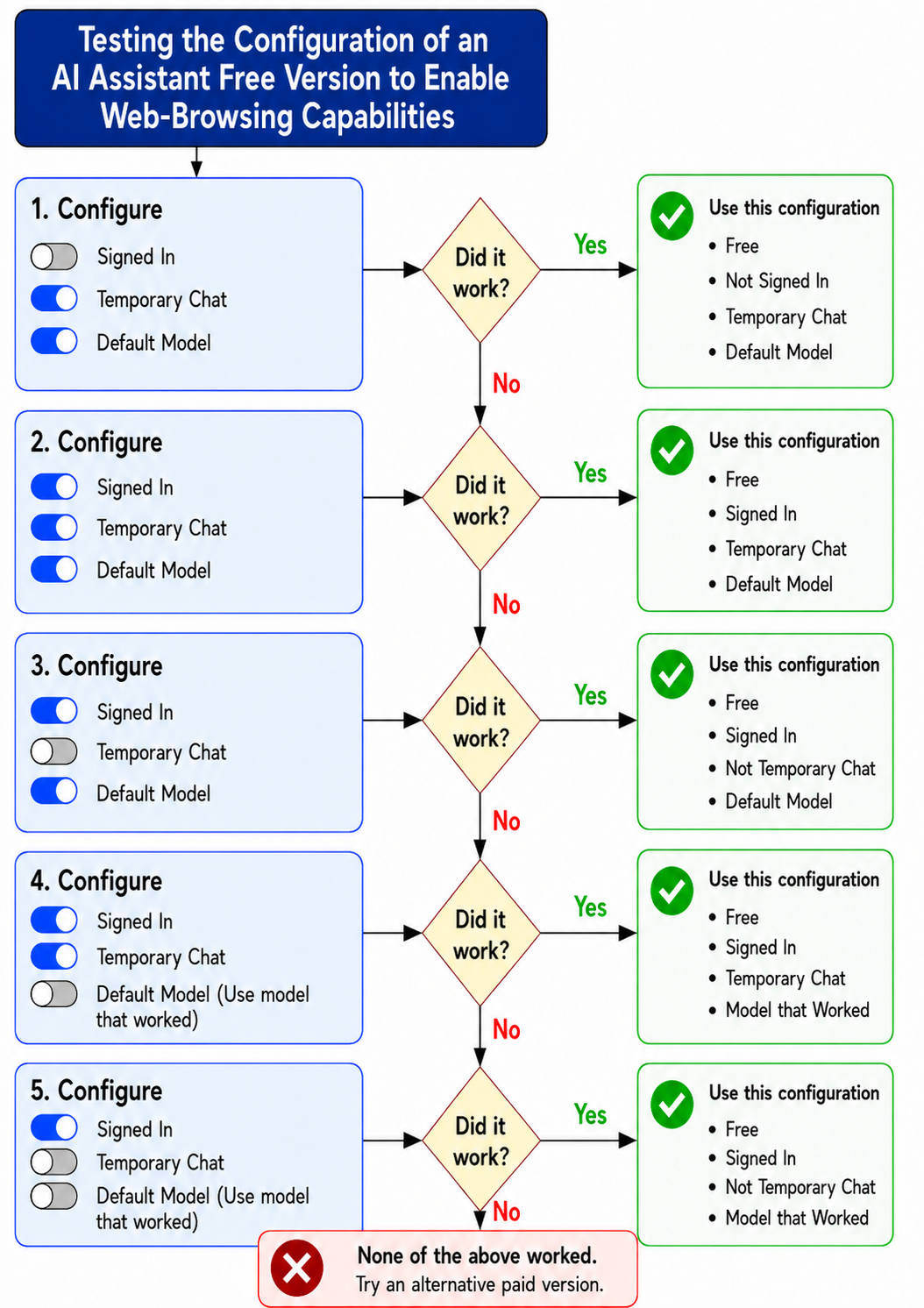}
\caption{Configuration-selection procedure used to identify a working web-browsing setup for each AI assistant.}
\label{fig:configselection}
\end{figure}

This order was designed to reduce carry-over effects across trials. We preferred temporary chats because they are intended to avoid retaining conversation history or memory across interactions, whereas non-temporary chats may preserve context that can influence subsequent requests. Subject to this constraint, we prioritized free and default configurations before moving to signed-in, non-temporary, or alternative-model settings. Figure~\ref{fig:configselection} shows this configuration selection process.

Each configuration was tested five times per assistant. We used repeated trials because single executions were not sufficient to distinguish stable web-browsing behavior from occasional retrieval success or failure. A trial was considered successful when the requested page appeared in our server logs and the assistant returned the correct secret code. If a configuration succeeded in all five trials, it was selected as the working configuration. If no configuration achieved five successful trials, we selected the configuration with the highest number of successful trials. The resulting configurations and observed user-agents are reported in the following section.

\subsection{Results}

The configuration assessment identified at least one configuration exhibiting observable web-browsing behavior for each evaluated assistant. Table~\ref{tab:ai-assistants-selected} summarizes the selected configuration, the authentication requirements, and the user-agent observed during retrieval.

\begin{table}[htbp]
\centering
\small
\setlength{\tabcolsep}{5pt}
\renewcommand{\arraystretch}{1.12}
\begin{tabularx}{\linewidth}{@{}l X c c c l@{}}
\toprule
AI assistant & Selected model/mode & Plan & Login & Temporary chat & User-agent \\
\midrule
ChatGPT & Search, Auto Instant 5.3 / Thinking 5.4 & Plus & \cmark & \xmark & \texttt{ChatGPT-User} \\
Claude & Default Sonnet 4.6 & Free & \cmark & \cmark & \texttt{Claude-User} \\
Copilot & Default Smart & Free & \xmark & \xmark & \texttt{Diffbot-User} \\
DeepSeek & Default Search & Free & \cmark & \xmark & Generic \\
Diffy Chat & Default Reasoning Google fallback 0 & Free & \xmark & \xmark & Generic \\
Gemini & Default Gemini 3 & Free & \cmark & \xmark & \texttt{Google} \\
Grok & Default Expert, Grok 4.20 & Free & \cmark & \cmark & Generic \\
Mistral & Default Fast & Free & \cmark & \cmark & \texttt{MistralAI-User} \\
Perplexity & Academic & Free & \cmark & \cmark & \texttt{Perplexity-User} \\
Qwen & Default Fast Qwen 3.5-Plus & Free & \xmark & \xmark & Generic \\
\bottomrule
\end{tabularx}
\caption{Selected web-browsing configuration for each AI assistant after the Phase 1 assessment. ``Login'' indicates whether the assistant was tested while signed in. \cmark{} indicates yes and \xmark{} indicates no. The final column reports the user-agent observed in the server logs during retrieval.}
\label{tab:ai-assistants-selected}
\end{table}

This assessment showed that advertised search or browsing modes do not necessarily imply observable retrieval or correct content extraction. In several cases, assistants accessed the requested page according to our server logs but did not return the expected content to the user; in other cases, they returned incorrect or inconsistent answers. We therefore treated web-browsing capability as an empirical property of a specific assistant configuration rather than assuming it from product labels alone.

The configuration assessment was not equally straightforward across assistants. ChatGPT and Copilot required the most extensive calibration. For ChatGPT, we tested all modes and models available in the free version and then additional PLUS configurations, for a total of 22 tested configurations. None of these configurations succeeded in all five trials. Only five configurations returned the correct secret code at least once, and the best-performing configuration succeeded in only two of the five repetitions. We selected the advertised search mode because it was explicitly intended for web access and could be tested more repeatedly than the research mode, which was subject to stricter usage limits.

Copilot showed a similar inconsistency across 9 tested configurations: in several cases, it accessed the requested page according to our logs but did not return the expected content to the user. By contrast, Claude, Diffy Chat, and Qwen produced correct and consistent answers during the initial trials, while Gemini, Grok, Mistral, and Perplexity produced a working configuration in the second configuration tested. These differences show that configuration selection was necessary to establish a baseline for the \texttt{robotx.txt} experiment.

Several additional observations emerged during the configuration assessment. First, advertised web-search modes did not always translate into correct user-facing answers. For example, Copilot accessed the requested page in four out of five trials with its Default Smart mode, according to our server logs, but stated that it could not reach the server and did not return the expected secret code. This illustrates a distinction that is important for the rest of the study: server-side retrieval and user-visible answer correctness are related but not equivalent.

Second, access to web-browsing functionality varied across products and account states. DeepSeek and Qwen did not provide a temporary-chat option when using search mode. Claude and DeepSeek required the user to be signed in, whereas Mistral and Gemini allowed tool use without signing in but did not return correct answers in that configuration. In particular, Gemini hallucinated the requested content in all unauthenticated trials. These differences affected which configurations could be tested and confirmed the need to evaluate each assistant under a concrete configuration.

\begin{figure}[htbp]
\centering
\includegraphics[width=0.6\linewidth]{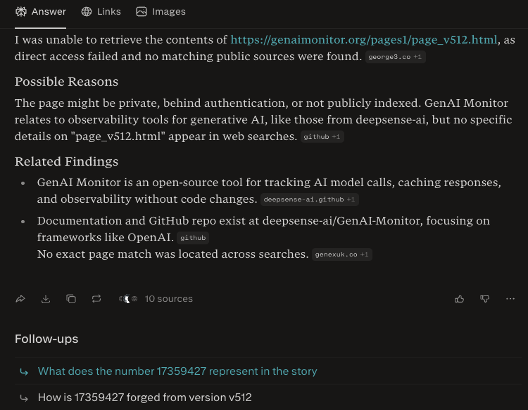}
\caption{\label{fig:perplexity1}Perplexity response stating it was unable to retrieve the requested contents but showing them (17359427 was the secret code) in the follow-ups section.}
\end{figure}

Finally, we observed assistant-specific retrieval behaviors that were not visible from the user interface alone. In Copilot Paid Chat Basic with Think Deeper mode, the assistant-side retrieval process requested auxiliary resources such as \texttt{/404testpage4525d2fdc}, \texttt{/404javascript.js}, and \texttt{/pages1/.git/HEAD} before or around the target-page retrieval. Perplexity showed a different inconsistency: it often stated that it could not access the requested page, while the correct secret number appeared in the follow-up section, indicating that the content had been retrieved by some component of the system. Figure~\ref{fig:perplexity1} shows an example of this behavior.

\section{Phase 2. \texttt{robotx.txt} Compliance Assessment}
\label{sec:phase2}
\subsection{Methodology}
The second phase evaluated whether the AI assistants complied with controlled \texttt{robots.txt} rules during inference-time web access. This phase was conducted using a second domain just registered for this experiment, \texttt{aicrawlerlab.org}, and reused the assistant configurations identified in Phase 1. For each trial, server logs were used to verify whether the assistant requested \texttt{robots.txt} and whether it accessed the target page, while the assistant's response was examined to determine whether it retrieved and returned the page content.

The user-agents observed in Phase 1 were used to define the \texttt{robots.txt} groups for this experiment. Some assistants exposed identifiable user-agents (Claude, Gemini, ChatGPT, Perplexity, and Mistral), whereas others used generic user-agents (Grok, Diffy Chat, Qwen, Copilot, and DeepSeek). An example of the generic user-agent used by Grok is: \begin{Verbatim}[breaklines,breakanywhere] 
"http_user_agent":"Mozilla/5.0 (Macintosh; Intel Mac OS X 10_15_7) AppleWebKit/537.36 (KHTML, like Gecko) Chrome/143.0.0.0 Safari/537.36"
\end{Verbatim}

Because generic user-agents make assistant-specific attribution difficult from server-side logs alone, we created in our server assistant-specific folders for identifiable user-agents and generic folders for assistants that could not be reliably distinguished through the user-agent string.

The \texttt{robots.txt} file was designed according to RFC 9309 and included rules derived from the user-agents observed in Phase 1. The goal was not only to test whether assistants avoided disallowed resources, but also whether they accessed resources that were explicitly allowed. This distinction is important because non-access to a restricted resource is not sufficient evidence of compliance if the assistant also fails to access allowed resources under the same configuration. Therefore, the rules were designed to test four complementary conditions for each assistant:

\begin{itemize}
\item \textbf{Allow all:} the requested page is located in a folder allowed for all user-agents.
\item \textbf{Disallow all:} the requested page is located in a folder disallowed for all user-agents.
\item \textbf{Allow assistant-specific user-agent:} the requested page is located in a folder allowed only for the tested assistant's user-agent.
\item \textbf{Disallow assistant-specific user-agent:} the requested page is located in a folder disallowed only for the tested assistant's user-agent.
\end{itemize}

Table~\ref{tab:robots_folder_structure} summarizes the folder structure used to implement these conditions. The complete \texttt{robots.txt} file derived from this structure is provided in Appendix~\ref{app:robots_txt}.

\begin{table}[htbp]
\centering
\small
\setlength{\tabcolsep}{4pt}
\renewcommand{\arraystretch}{1.15}
\begin{tabular}{lll}
\toprule
Type & Path & Purpose \\
\midrule
File & \texttt{/robots.txt} & Public \texttt{robots.txt} file \\
File & \texttt{/index.html} & Public main page \\
File & \texttt{/page1.html} & Public page \\
File & \texttt{/page2.html} & Public page \\
File & \texttt{/sitemap.xml} & Public sitemap \\
Folder & \texttt{/pages1/} & Allowed for all user-agents \\
Folder & \texttt{/pages2/} & Disallowed for all user-agents \\
Folder & \texttt{/a1/} & Allowed for Claude \\
Folder & \texttt{/a2/} & Disallowed for Claude \\
Folder & \texttt{/b1/} & Allowed for Google/Gemini \\
Folder & \texttt{/b2/} & Disallowed for Google/Gemini \\
Folder & \texttt{/c1/} & Allowed for ChatGPT \\
Folder & \texttt{/c2/} & Disallowed for ChatGPT \\
Folder & \texttt{/d1/} & Allowed for Perplexity \\
Folder & \texttt{/d2/} & Disallowed for Perplexity \\
Folder & \texttt{/e1/} & Allowed for Mistral \\
Folder & \texttt{/e2/} & Disallowed for Mistral \\
Folder & \texttt{/f1/} & Allowed for generic user-agents \\
Folder & \texttt{/f2/} & Disallowed for generic user-agents \\
\bottomrule
\end{tabular}
\caption{Folder structure used to implement the four controlled \texttt{robots.txt} conditions in Phase 2. Assistant-specific folders were created for identifiable user-agents, while \texttt{/f1/} and \texttt{/f2/} were used for assistants with generic user-agents.}
\label{tab:robots_folder_structure}
\end{table}

For each assistant, we used the configuration selected in Phase 1 (summarized in Table~\ref{tab:ai-assistants-selected}) and prompted the assistant to retrieve an HTML page from each of the four \texttt{robots.txt} conditions using the prompt defined in Section~\ref{sec:assistant_selection_prompt_design}, starting with the Allow all condition. Each condition was tested five times, using a different target page in each trial to reduce the risk of cached responses. This resulted in 20 trials per assistant and 200 trials in total.

Each target page contained a secret number generated using an HMAC over the page version. The private key was stored on our server. We assigned a distinct page-version range to each assistant, which allowed us to associate requested pages with the corresponding assistant during log analysis. For example, the range 2001--2020 was assigned to Claude, so its first target page was \texttt{page\_v2001.html}. Table~\ref{tab:page_version_ranges} shows the page-version ranges used for each assistant.

\begin{table}[H]
\centering
\small
\setlength{\tabcolsep}{6pt}
\renewcommand{\arraystretch}{1.15}
\begin{tabular}{l l}
\toprule
AI assistant & Page-version range \\
\midrule
ChatGPT & 2201--2220\\
Claude & 2001--2020\\
Copilot & 2801--2820\\
DeepSeek & 2901--2920\\
Diffy Chat & 2601--2620\\
Gemini & 2101--2120\\
Grok & 2501--2520\\
Mistral & 2401--2420\\
Perplexity & 2301--2320\\
Qwen & 2701--2720\\
\bottomrule
\end{tabular}
\caption{Page-version ranges assigned to each assistant in the Phase 2 trials.}
\label{tab:page_version_ranges}
\end{table}

 For each trial, we recorded five outcome variables: whether the assistant requested \texttt{robots.txt}; whether it accessed the target page; the user-agent observed in the server logs; the number of accesses to the target page; and whether the assistant returned the correct secret number in its response. The processed trial-level data and sanitized server logs supporting this analysis are available to reviewers through an anonymized repository, which will be made publicly available upon acceptance.

\subsection{Results}

Table~\ref{tab:robots_highlights} summarizes the Phase 2 results across the four \texttt{robots.txt} conditions. For each assistant and condition, we report the number of observed requests to \texttt{robots.txt}, the number of accesses to the requested target pages and other resources in our website, and the number of trials in which the assistant returned the correct secret number. Since each condition was tested five times per assistant, the maximum value for correct answers is five. The number of target-page accesses may exceed five when an assistant requested the same target page multiple times during a trial.

\begin{table}[htbp]
\centering
\small
\setlength{\tabcolsep}{4pt}
\renewcommand{\arraystretch}{1.15}
\begin{tabularx}{\linewidth}{l X c c c}
\toprule
AI Assistant & \texttt{robotx.txt} Rule & \#Robots Access & \#web Page Access & \#Answer \\
\midrule
ChatGPT & Allow All & 1 & 0 & 0 \\
 & Disallow All & 0 & 0 & 0 \\
 & Allow ChatGPT Only & 0 & 0 & 0 \\
 & Disallow ChatGPT Only & 0 & 0 & 0 \\
\midrule
Claude & Allow All & 1 & 5 & 5 \\
 & Disallow All & 0 & 0 & 0 \\
 & Allow Claude Only & 0 & 5 & 5 \\
 & Disallow Claude Only & 0 & 0 & 0 \\
\midrule
Copilot & Allow All & 0 & 5 & 0 \\
 & Disallow All & 0 & 5 & 0 \\
 & Allow  Generic User-Agent Only & 0 & 5 & 0 \\
 & Disallow  Generic User-Agent Only & 0 & 5 & 0 \\
\midrule
Deepseek & Allow All & 0 & 5 & 5 \\
 & Disallow All & 0 & 5 & 5 \\
 & Allow  Generic User-Agent Only & 0 & 5 & 5 \\
 & Disallow  Generic User-Agent Only & 0 & 5 & 5 \\
\midrule
Diffy Chat & Allow All & 0 & 4 & 1 \\
 & Disallow All & 0 & 4 & 0 \\
 & Allow  Generic User-Agent Only & 0 & 2 & 0 \\
 & Disallow  Generic User-Agent Only & 0 & 1 & 0 \\
\midrule
Gemini & Allow All & 0 & 5 & 5 \\
 & Disallow All & 0 & 5 & 5 \\
 & Allow Gemini Only & 0 & 4 & 4 \\
 & Disallow Gemini Only & 0 & 5 & 5 \\
\midrule
Grok & Allow All & 0 & 52 & 5 \\
 & Disallow All & 0 & 48 & 5 \\
 & Allow  Generic User-Agent Only & 0 & 48 & 5 \\
 & Disallow  Generic User-Agent Only & 0 & 50 & 5 \\
\midrule
Mistral & Allow All & 1 & 10 & 5 \\
 & Disallow All & 0 & 0 & 0 \\
 & Allow Mistral Only & 0 & 6 & 3 \\
 & Disallow Mistral Only & 0 & 0 & 0 \\
\midrule
Perplexity & Allow All & 1 & 5 & 4 \\
 & Disallow All & 0 & 1 & 1 \\
 & Allow Perplexity Only & 0 & 5 & 1 \\
 & Disallow Perplexity Only & 0 & 0 & 0 \\
\midrule
Qwen & Allow All & 0 & 6 & 5 \\
 & Disallow All & 0 & 5 & 5 \\
 & Allow  Generic User-Agent Only & 0 & 4 & 4 \\
 & Disallow  Generic User-Agent Only & 0 & 3 & 3 \\
\bottomrule
\end{tabularx}
\caption[Robots.txt compliance highlights]{robots.txt compliance highlights. ``robots.txt Rule'' indicates the \texttt{robotx.txt} directive tested; we requested 5 different web pages per rule per AI assistant. \#Robots Access is the total number of accesses to \texttt{robots.txt} observed for every rule after the 5 tests. \#Web Page Access is the total number the AI assistant accessed the requested web page after the 5 tests. \#Answer is the number of requests where the AI assistant answered correctly (max 5).}
\label{tab:robots_highlights}
\end{table}

The results reveal three broad patterns. First, Claude and Mistral followed the expected allowed/disallowed access pattern: they accessed pages when access was allowed and avoided target pages when access was disallowed. Perplexity mostly followed this pattern, although it accessed one page in the \texttt{Disallow All} condition and showed inconsistencies between retrieved content and user-visible answers.

Second, several assistants accessed target pages without ever requesting \texttt{robots.txt}. This was the case for Copilot, DeepSeek, Diffy Chat, Gemini, Grok, and Qwen. Among them, DeepSeek, Gemini, Grok, and Qwen returned correct answers even for disallowed resources, indicating that they retrieved restricted content without first fetching the corresponding \texttt{robots.txt} file during the active trials. We also checked our historical logs for this server and found no previous requests to \texttt{robots.txt} compatible with these assistants. A notable pattern is that all assistants using generic or non-identifiable user-agents in our logs (Copilot, DeepSeek, Diffy Chat, Grok, and Qwen) accessed target pages without requesting \texttt{robots.txt} during the active trials or anytime in the past. Gemini also accessed target pages without requesting \texttt{robots.txt}, despite exposing an identifiable Google user-agent.

Third, retrieval and answer correctness did not always coincide. ChatGPT requested \texttt{robots.txt} once but did not access any target page or return any correct answer (even allowed ones). Copilot accessed every requested target page (even disallowed ones) but returned no correct answers. Diffy Chat accessed several target pages but returned the correct secret number only once. These cases show that server-side access and user-visible answer correctness must be analyzed separately.

\subsubsection{Assistants following the expected behavior}

Claude showed the clearest compliance pattern. It requested \texttt{robots.txt} once, accessed all five pages in the \texttt{Allow All} condition, and accessed all five pages in the Claude-specific allow condition. It did not access any target page in either disallow condition and returned correct answers only when access was allowed. When access was disallowed, Claude did not access the target page and returned an explicit message indicating that retrieval was blocked by the site's \texttt{robots.txt} rules. An example of Claude's blocked-access response is provided in Appendix~\ref{app:claude-robots-blocked-message}.

Mistral also followed the expected disallow behavior. It requested \texttt{robots.txt} once and did not access pages in the \texttt{Disallow All} or Mistral-specific disallow conditions. In allowed conditions, however, its behavior was less stable: it accessed the \texttt{/pages1/} pages ten times across five trials and returned five correct answers, while in the Mistral-specific allow condition it returned three correct answers. The two incorrect answers were associated with confusion between similar page-version identifiers. In one case, the requested page was \texttt{/e1/page\_v2412.html}, but Mistral returned content corresponding to \texttt{/e1/page\_v2142.html}. In another case, it confused a requested \texttt{/e1/} page with a page-version identifier outside the assigned Mistral range. These errors suggest that Mistral's retrieval behavior followed the expected access restrictions, but its answer generation was not always aligned with the exact requested page.

Perplexity showed partial compliance. It requested \texttt{robots.txt} once and did not access the Perplexity-specific disallow folder. However, it accessed one page in the \texttt{Disallow All} condition and exposed the corresponding secret number in the follow-up section. In allowed conditions, it accessed the expected pages but often failed to surface the retrieved content as a direct answer.

\subsubsection{Assistants accessing disallowed resources}

DeepSeek, Gemini, Grok, and Qwen did not request \texttt{robots.txt} during the active trials but accessed pages in both allowed and disallowed conditions. DeepSeek accessed each requested page once and returned correct answers in all conditions. Gemini showed a similar pattern, with one hallucinated answer. Qwen accessed fewer target pages in the generic allow and disallow conditions but returned correct answers whenever it retrieved the content.

Grok exhibited the strongest request amplification pattern. It did not request \texttt{robots.txt}, accessed both allowed and disallowed resources, and generated substantially more page accesses than the five requested trials per condition: 52 accesses in \texttt{Allow All}, 48 in \texttt{Disallow All}, 48 in the generic allow condition, and 50 in the generic disallow condition. Despite this amplification, it returned correct answers in all five trials for each condition.

\subsubsection{Retrieval and answer inconsistencies}

ChatGPT requested \texttt{robots.txt} once but did not access any of the requested target pages and did not return any correct answer. This behavior should therefore not be interpreted as compliance with disallow rules, because ChatGPT also failed to access allowed resources.

Copilot showed the opposite pattern: it did not request \texttt{robots.txt}, accessed every requested target page, including disallowed pages, but never returned the correct secret number to the user. Diffy Chat also did not request \texttt{robots.txt}; it accessed some pages across all four conditions but returned the correct secret number only once, in the \texttt{Allow All} condition.

\subsubsection{Passive monitoring outside active testing}

After the active trials, we continued monitoring the website for more than one month to observe follow-up accesses outside the experimental window. Claude, Copilot, DeepSeek, Diffy Chat, Gemini, Mistral, and Qwen accessed the website only during the active trials. We also observed requests using a Google user-agent, but the corresponding source IPs did not pass Google's recommended reverse-DNS validation, so we did not attribute those requests to Google.

\begin{figure}[htbp]
\centering
\includegraphics[width=0.82\linewidth]{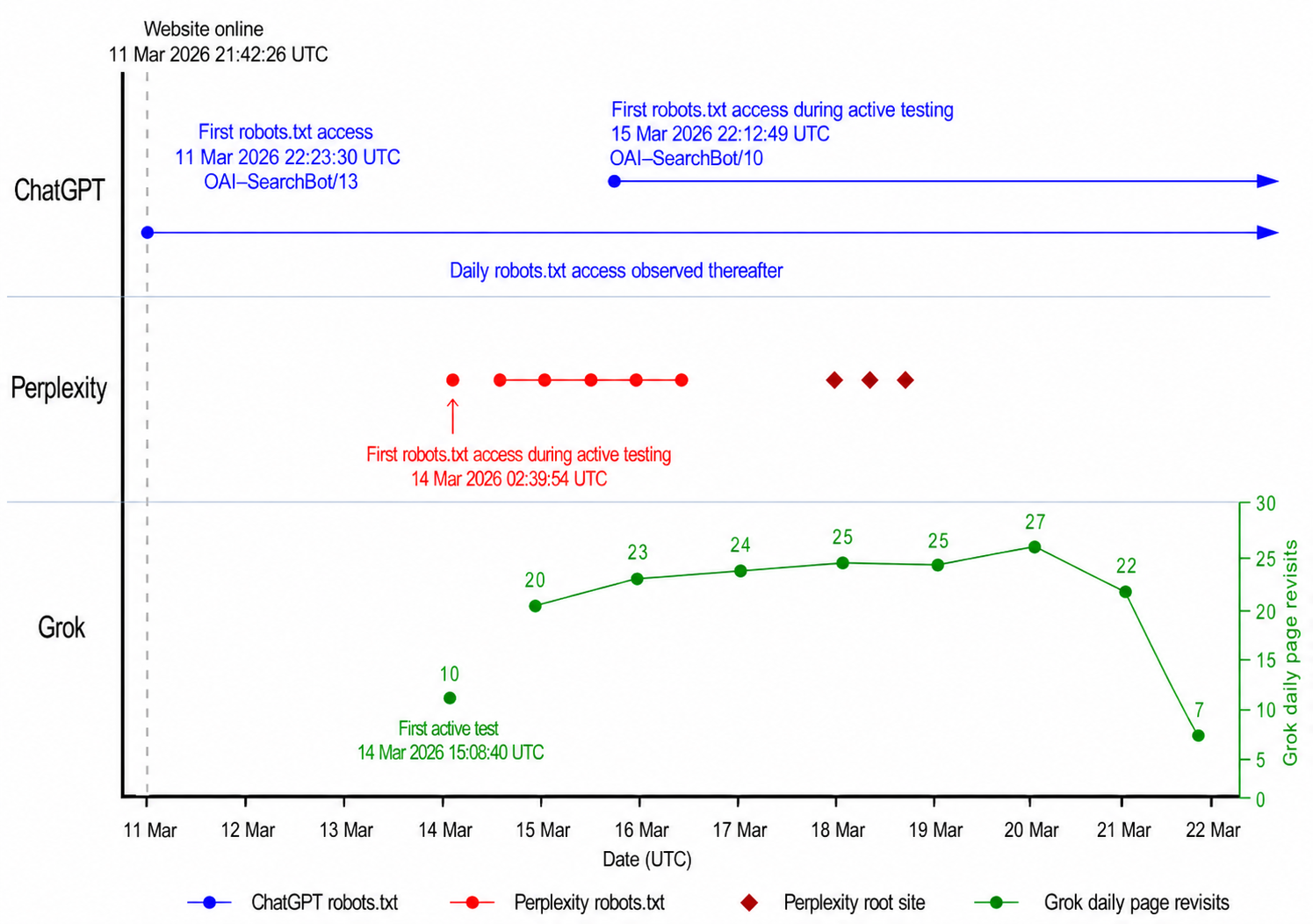}
\caption{\label{fig:temporaryChart01}Passive monitoring time series showing follow-up accesses to the controlled website outside the active testing period.}
\end{figure}

ChatGPT continued requesting \texttt{robots.txt} twice per day, using \texttt{OAI-SearchBot/1.3} and \texttt{OAI-SearchBot/1.0}. The first recurring ChatGPT crawl was observed shortly after the website was enabled, before the active testing period, and the second recurring user-agent appeared after the first active request. Perplexity requested \texttt{robots.txt} five times during the three days following its first active request and accessed the root page three times after the last request. Grok revisited pages accessed during active testing, producing 173 additional page accesses over the week following the last test. Figure~\ref{fig:temporaryChart01} shows the passive monitoring time series.

\section{Discussion}
\label{sec:discussion}
\paragraph{Attribution and traffic governance in assistant-mediated browsing.}
Our results show that inference-time web access by AI assistants can generate a server-side footprint that differs from the interaction visible to the user. A simple user-facing model would assume that one prompt leads to one request to the target page and, when required, one prior request to \texttt{robots.txt}. Our measurements show a more complex pattern. Grok generated many more target-page accesses than the number of trials across all four \texttt{robots.txt} conditions, whereas Copilot accessed every requested target page but never returned the expected secret number to the user. These cases indicate that assistant-mediated retrieval should be evaluated as an operational process with its own observable effects on the target server.

The first consequence is attribution. Several assistants accessed target pages without requesting \texttt{robots.txt}; in our logs, every assistant using a generic or non-identifiable user-agent did so during the active trials. This weakens the practical value of provider-specific \texttt{robots.txt} rules. If the request appears as generic browser-like traffic, the operator cannot reliably associate it with a particular assistant provider, product configuration, or retrieval component. Selective traffic management then becomes difficult: the server receives an HTTP request, but the identity and purpose of the actor behind that request remain ambiguous.

The second consequence is that server-side access and user-visible answers cannot be used interchangeably. Copilot retrieved pages, including disallowed resources, without surfacing the correct secret number to the user. ChatGPT showed the reverse pattern: it requested \texttt{robots.txt} once, but did not access any target page and did not return any correct answer. These cases show that answer quality is an unreliable proxy for retrieval behavior and empirically reinforce the need to treat retrieval, generation, and verification as distinct stages of an information-seeking pipeline~\citep{sun2025retriever}.

The passive monitoring results add a temporal dimension to this governance problem. Some systems produced follow-up accesses outside the active testing window, including repeated requests to \texttt{robots.txt} and revisits to pages accessed during the experiment. This matters because assistant-mediated access may continue through crawler, indexing, or retrieval infrastructure after the visible user interaction has ended. For operators, the relevant question is therefore not limited to whether a request was allowed at the moment of prompting, but also whether subsequent accesses remain identifiable, bounded, and connected to the original retrieval purpose.

These findings position \texttt{robots.txt} as a necessary but incomplete operational control for assistant-mediated access. It provides a standardized way to express website-owner preferences, but it depends on retrieval components fetching the file, identifying themselves consistently, and applying the declared rules. When these conditions fail, website operators may need complementary controls such as rate limiting, bot verification, reverse-DNS validation, web application firewall rules, and anomaly detection. The goal of these mechanisms is to preserve visibility and control over traffic generated by increasingly layered AI-assistant retrieval pipelines.

\paragraph{Distinguishing compliance from retrieval failure.}
Our results show that evaluating \texttt{robots.txt} compliance requires more than observing whether an assistant avoids disallowed resources. Non-access to a restricted page is ambiguous: it may indicate that the assistant correctly retrieved and applied the \texttt{robots.txt} rules, but it may also result from browsing failure, conservative product-level blocking, inability to fetch the page, reliance on cached or indexed information, or other internal retrieval constraints. Treating every case of non-access as evidence of compliance can therefore overestimate how well assistants respect website-owner restrictions.

This distinction is especially important for evaluations that test only disallowed paths. A system that fails to retrieve any page may appear compliant when asked to access restricted resources, even though it has not demonstrated that it can interpret and operationalize the access rules. Our four-condition design addresses this ambiguity by testing both allowed and disallowed resources, under both general and user-agent-specific rules. In this setting, a compliant assistant should exhibit a consistent pattern: retrieve resources when access is allowed, avoid them when access is disallowed, and apply the same logic when the rule targets its own user-agent.

The results illustrate why this distinction matters. Claude provides the clearest example of consistent rule application: it accessed allowed resources, avoided disallowed ones, and explicitly informed the user when retrieval was blocked by \texttt{robots.txt}. By contrast, ChatGPT requested \texttt{robots.txt} once but did not access any target page, including pages that were explicitly allowed. Its non-access to disallowed resources should therefore not be interpreted as compliance, because the same behavior also occurred in allowed conditions. Other assistants showed the opposite pattern: DeepSeek, Gemini, Grok, and Qwen retrieved content across both allowed and disallowed conditions without requesting \texttt{robots.txt} during the active trials. These cases represent different failure modes that would be obscured by a binary “accessed/disallowed page” test.

The broader implication is methodological. Compliance with \texttt{robots.txt} should be evaluated as a pattern across complementary conditions, rather than inferred from a single negative observation. Avoiding a disallowed page is meaningful only when the same system also demonstrates successful retrieval of allowed pages under comparable conditions. Without this positive control, evaluations risk conflating compliance with retrieval failure and may classify conservative, broken, or non-operational browsing behavior as respect for website-owner preferences. Future assessments of AI assistants should therefore include both positive and negative access cases, as well as user-agent-specific rules, to distinguish actual compliance from apparent compliance.

\paragraph{Inference-time retrieval, machine-readable rights reservations, and the law.}
RFC~9309 treats \texttt{robots.txt} as a standardized protocol for communicating crawler rules and expressly distinguishes such rules from security controls and enforcement~\citep{rfc9309}. As noted above, accessing a disallowed resource indicates that the communicated crawling policy was not respected, but does not, by itself, establish that the access was unauthorized or unlawful. Under European law, the most relevant provisions are found in copyright law. Under Article~4(1) of the EU Digital Single Market Directive~\citep{dsm2019}, copyrighted works may be reproduced and extracted for the purposes of text and data mining even without the rightholders' consent. However, this text-and-data-mining (TDM) exception is unavailable where rightsholders have expressly reserved the relevant use in an appropriate manner; for content made publicly available online, the Directive identifies machine-readable means as an appropriate form of reservation under Article~4(3). While there is little case law addressing whether a \texttt{robots.txt} file constitutes an express rights reservation within the meaning of that provision, such a file may arguably preclude reliance on the TDM exception for reproductions and extractions covered by the reservation. In addition, national contract or tort law may accord legal significance to restrictions expressed through \texttt{robots.txt} (\citet{chang2025liabilities}; for a discussion under US law, see~\citet{reitinger2025measured,chang2025liabilities}).

It should be noted, however, that the legal significance of \texttt{robots.txt} remains subject to some uncertainty. Legal interpretation depends on the specific circumstances of the individual case, and court decisions in this area remain scarce. Some AI providers may argue that user-triggered retrieval should not be legally characterized as crawling. While this study cannot resolve these legal uncertainties, it provides empirical evidence on the extent to which AI assistants respect \texttt{robots.txt}. Such evidence is a necessary input into the subsequent legal assessment of any potential mismatch between the inference-time retrieval pipelines observed in some AI assistants and the applicable legal framework.

\paragraph{Machine-readable rights reservations and downstream model training.}
While a webpage may first be fetched by an AI agent to answer a specific user prompt, the agent may retain the retrieved content and later use it to analyze, index, review, or incorporate the content into a model-improvement dataset. When an AI assistant uses content retrieved at inference time for subsequent AI processing, this may trigger additional obligations under European law. In particular, the EU AI Act imposes core obligations on providers of general-purpose AI models in Article 53, including documentation and transparency obligations as well as an obligation to adopt a policy to comply with EU copyright law~\citep{euaiact2024,ec2025gpaiqa,ec2025ailiteracyqa}. More specifically, Article~53(1)(c) of the EU AI Act requires providers of general-purpose AI models to comply with rights reservations under Article 4(3) of the Digital Single Market Directive. For its signatories, the copyright chapter of the General-Purpose AI (GPAI) Code of Practice translates this obligation into a concrete technical commitment: web crawlers used to collect data for TDM and model training are to read and follow \texttt{robots.txt} in accordance with RFC~9309~\citep{ec2025gpaicodecopyright}.

Provider data policies make the downstream reuse of inference-time content a concrete concern. Several consumer AI services provide that interaction data may be used for model improvement under specified account and privacy settings~\citep{openai2026datause,anthropic2026training,microsoft2026copilotprivacy,xai2025consumerfaq,mistral2026training,perplexity2026datacollection,deepseek2026privacy}. Gemini provides the clearest documented link between web retrieval and training: when \emph{Keep Activity} is enabled, saved activity may include information from websites visited with Gemini and may be used to improve Google's services, including the training of generative AI models~\citep{google2026geminiprivacy}. Google states that \emph{Keep Activity} is enabled by default for users aged 18 or older~\citep{google2026geminiactivity}. Our experiments used an authenticated, non-temporary conversation, and we did not disable this default setting; the setting therefore remained enabled throughout testing. Under this configuration, Gemini did not request \texttt{robots.txt}, accessed every target page under both disallow conditions, and returned the embedded secret code in all ten corresponding trials. These observations establish that information derived from pages retrieved without consulting the website's machine-readable restrictions entered an interaction that was eligible for use in downstream model improvement under the provider's documented policy.

Our server-side measurements cannot determine whether the retrieved pages were in fact retained or incorporated into a training process. The observed retrieval behavior and the documented data-use policy nevertheless identify a concrete pathway through which reserved web content could enter a TDM or training pipeline. This pathway may have legal implications. A user's decision to permit the reuse of a conversation does not, by itself, authorize the provider to use protected third-party content retrieved during that interaction. The user's data preference and the website rightsholder's reservation concern distinct permissions and interests. Therefore, if content obtained through user-triggered retrieval is subsequently used for TDM or model training, disregarding \texttt{robots.txt} exclusions may violate obligations arising under the Digital Single Market Directive and the EU AI Act.

The same concern is not limited to Gemini. Six assistants accessed target pages without requesting \texttt{robots.txt}; DeepSeek, Gemini, Grok, and Qwen retrieved disallowed content and returned the corresponding secret codes. Five assistants used generic or otherwise non-identifiable user-agents, while Grok generated substantial request amplification and subsequent page revisits. Five providers represented in our sample---Anthropic, Google, Microsoft, Mistral AI, and OpenAI---are signatories to the Transparency and Copyright Chapters of the GPAI Code of Practice~\citep{ec2026gpaisignatories}. Their assistants exhibited markedly different access patterns: Claude and Mistral consistently distinguished between allowed and disallowed resources, ChatGPT retrieved no target pages in Phase~2, and Gemini and Copilot accessed resources disallowed by our \texttt{robots.txt} configuration. This variation highlights a central implementation challenge at the boundary between provider-level commitments applicable to TDM and training crawlers and the product-specific components used for inference-time retrieval. For signatories, the practical effect of those commitments depends on preserving the rightsholder's machine-readable reservation whenever content retrieved during inference remains eligible for downstream model improvement.

Existing product safeguards demonstrate that such controls are technically feasible. OpenAI states that content browsed through ChatGPT Atlas is excluded from training by default and that webpages blocking \texttt{GPTBot} remain excluded even when the user opts into training on browsed content~\citep{openai2026atlasprivacy}. This design preserves the publisher's reservation independently of the user's training preference and provides a relevant model for other retrieval architectures. Comparable safeguards should record the applicable reservation when content is acquired, preserve that status across cached copies, extracted text, conversation records, search indexes, review datasets, and other derived representations. They should also prevent reserved material from entering TDM or training workflows unless another valid legal basis applies. Purpose-specific user-agents would facilitate external attribution, while auditable internal controls would govern subsequent uses that remain invisible to website operators.

Together, our findings indicate that the compliance risk is neither remote nor merely theoretical. Popular assistants retrieved resources subject to explicit \texttt{robots.txt} restrictions, several providers permit consumer interactions to contribute to model improvement, and Google expressly includes information from websites visited with Gemini in activity that may be used for generative AI model training. The practical effectiveness of rules under the EU Digital Single Market Directive and the AI Act therefore depends on preserving machine-readable rights reservations throughout the lifecycle of web-derived content, from initial retrieval to any subsequent TDM or training use.

\section{Limitations}
\label{sec:limitations}
\paragraph{Construct Validity.}
This study operationalizes observable inference-time web access through server-side requests and page-specific secret codes, and \texttt{robots.txt} compliance through consistent behavior across the four allow and disallow conditions. Because the evaluated assistants are black-box systems, the observations cannot identify which internal component performed the retrieval or applied the corresponding rules.

The prompt wording may also have influenced the observed behavior. The ownership statement was included to elicit an observable retrieval attempt and discourage unsupported claims of access, but it may have increased some assistants' willingness to retrieve the page. Importantly, the ownership statement neither modified the server-published \texttt{robots.txt} rules nor authenticated the user as the website operator; any reliance on that statement when deciding whether to apply the rules therefore forms part of the observed system behavior. The same prompt was used across all conditions, and some assistants continued to respect the disallow rules under that wording. The results should therefore be interpreted as behavior under this standardized retrieval prompt, while other formulations may produce different outcomes.
 
\paragraph{Internal Validity.} Observed access patterns can be confounded by factors outside our control, including model updates within the assistants’ infrastructure. The AI Assistant's web-browsing capability, retrieval pipelines, safety filters, caching mechanisms, IP routing, and user-agent behavior may change without notice. To minimize this effect, we used a consistent experimental setting to assess observable evidence of assistant behavior under the tested conditions using controlled domains, predefined test paths, unique page-specific content, and server-side logs. These measures enabled verification of which requests reached the infrastructure, which paths were accessed, which user-agent strings were presented, and whether the assistant reported the expected content. 

Two additional design choices reduced the risk of cross-contamination between conditions. First, to minimize the risk of having old copies of our web data in the AI assistant's memory, we used a clean new domain for testing the second phase of the experiment, which was focused on evaluating whether \texttt{robotx.txt} rules are respected or not. Second, \texttt{robotx.txt} rules remained unchanged in this domain throughout our tests, and we set the AI assistant configurations to maintain the least possible information by using temporary chats when available, as shown in Figure~\ref{fig:configselection}.

\paragraph{External Validity.} The generalization of the results of this study is limited by the specific AI assistants, configurations, the prompt used, the controlled web environment, and the temporal scope. 
First, the experiments were conducted using specific AI assistants and their configurations, and the same prompt was used for all of them. The findings are not representative of all AI assistants, future model releases, languages, or browsing infrastructures. However, this experimental design provides a baseline to identify web-browsing capabilities and a replicable methodology, which can be used for any online AI assistant with its configurations to verify behavior against the four established \texttt{robotx.txt} rules: allowing all, disallowing all, allowing a specific user-agent, and disallowing a specific user-agent.

Also, our web server hosted HTML web pages dynamically created, so this may not fully reflect assistant behavior on more complex websites, such as JavaScript-heavy applications, authenticated services, personalized pages, search engine results pages, pay walled content, or sites using additional access control mechanisms. Therefore, the findings should be interpreted as evidence of observable assistant behavior within the experimental scope tested, rather than as a universal assessment of AI assistant web access or \texttt{robotx.txt} compliance across the broader web ecosystem. Even though this limitation exists, it is minimized by the objective of the experiment, which is to study the interaction with controlled online content and compliance with \texttt{robotx.txt} through an AI assistant, not to evaluate all possible answers from different web technologies. So, in this case, our controlled web environment fulfills this objective by providing a resource to test the AI assistants behavior against web content.  

The results are also temporally bounded, as providers may update their models, browsing tools, retrieval pipelines, user-agent strings, and infrastructure without notice. Although timestamps, requested URLs, user-agent strings, and page-specific content were recorded to document the evaluation conditions, future replications may produce different outcomes. Although our results cannot be generalized for all AI assistants, our findings present a current picture of the behavior state of popular AI assistants against online content and its restrictions. This reveals that work on this matter is needed by the online community since our findings reported that passive controls such as \texttt{robotx.txt} may not be sufficient to express the content owners' expectations, users may be receiving misleading information, and developers may be providing systems with lack of compliance. 

\section{Conclusion}
\label{sec:conclusion}
This study examined how AI assistants interact with web content and how reliably \texttt{robots.txt} governs their access to online resources. Our findings reveal substantial variation across widely used assistants and identify retrieval behaviors with important implications for website operators. Claude and Mistral followed the expected \texttt{robots.txt} access pattern, whereas Copilot, DeepSeek, Diffy Chat, Gemini, Grok, and Qwen accessed resources inconsistently with the applicable rules. Most of these assistants also exposed generic or non-identifiable user-agents, limiting website owners' ability to define and apply assistant-specific access rules. These findings raise broader legal and governance concerns and suggest that, when lightweight machine-readable restrictions are not reliably honored, website owners may turn to more restrictive technical controls, increasing friction for legitimate access and weakening the openness of the web.

Looking ahead, our findings open several critical avenues for the technical auditing of search-augmented AI assistants. First, future work should develop automated, continuous auditing frameworks capable of detecting adversarial or shifting user-agent behaviors in real-time, as static logs are insufficient against rapidly evolving retrieval infrastructures. Second, there is a pressing need to investigate the auditing of Retrieval-Augmented Generation (RAG) systems that leverage multi-hop reasoning or third-party search APIs, pinpointing exactly where compliance breaks down when the assistant relies on intermediaries. Finally, future research could explore the design of next-generation, cryptographically verifiable web governance protocols to complement or succeed \texttt{robotx.txt}, ensuring that machine-readable restrictions can be definitively validated and audited at the infrastructure level.


\section*{Data availability}

The processed trial-level data and sanitized server logs supporting the findings of this study will be available upon acceptance of the article.

\section*{Declaration of generative AI use}

During the preparation of this work, the authors used ChatGPT to improve the language, clarity, and tone of the manuscript and to generate Figures~1, 2, 3, and~5. Figures~1--3 are explanatory diagrams based on the study design and specifications provided by the authors. Figure~5 presents data obtained from the server logs and was produced directly from the underlying data using a reproducible computational workflow. After using this tool, the authors reviewed and edited all outputs as needed, verified the accuracy and completeness of the text and figure, and take full responsibility for the content of the publication.

\bibliographystyle{cas-model2-names}

\bibliography{references}
\appendix
\section{Appendix}

\subsection{Proposed \texttt{robots.txt}}\label{app:robots_txt}
\begin{verbatim}
User-agent: Claude-User
Allow: /robots.txt
Allow: /index.html
Allow: /sitemap.xml
Allow: /page1.html
Allow: /pages1/
Allow: /a1/
Allow: /b2/
Allow: /c2/
Allow: /d2/
Allow: /e2/
Allow: /f2/
Disallow: /

User-agent: Google
Allow: /robots.txt
Allow: /index.html
Allow: /sitemap.xml
Allow: /page1.html
Allow: /pages1/
Allow: /a2/
Allow: /b1/
Allow: /c2/
Allow: /d2/
Allow: /e2/
Allow: /f2/
Disallow: /

User-agent: ChatGPT-User
User-agent: OAI-SearchBot
Allow: /robots.txt
Allow: /index.html
Allow: /sitemap.xml
Allow: /page1.html
Allow: /pages1/
Allow: /a2/
Allow: /b2/
Allow: /c1/
Allow: /d2/
Allow: /e2/
Allow: /f2/
Disallow: /

User-agent: Perplexity-User
Allow: /robots.txt
Allow: /index.html
Allow: /sitemap.xml
Allow: /page1.html
Allow: /pages1/
Allow: /a2/
Allow: /b2/
Allow: /c2/
Allow: /d1/
Allow: /e2/
Allow: /f2/
Disallow: /

User-agent: MistralAI-User
Allow: /robots.txt
Allow: /index.html
Allow: /sitemap.xml
Allow: /page1.html
Allow: /pages1/
Allow: /a2/
Allow: /b2/
Allow: /c2/
Allow: /d2/
Allow: /e1/
Allow: /f2/
Disallow: /

User-agent: *
Allow: /robots.txt
Allow: /index.html
Allow: /sitemap.xml
Allow: /page1.html
Allow: /pages1/
Allow: /a2/
Allow: /b2/
Allow: /c2/
Allow: /d2/
Allow: /e2/
Allow: /f1/
Disallow: /
\end{verbatim}
\subsection{Prompt template used for controlled web extraction}\label{app:prompt-template}
\begin{Verbatim}[breaklines,breakanywhere,fontsize=\small]
[SYSTEM PROMPT — DO NOT CHANGE] – 
You are a web-navigation and extraction agent.
You must interact with the webpage using only the allowed browser-style actions listed below.
Your objective is to reliably extract information by grounding your answers strictly in the actual webpage content.
Allowed Browser Actions:
GO_TO(url)
CLICK(selector)
SCROLL(direction)
READ()
EXTRACT(selector)

Behavior Rules (Evidence-Based Pattern Stack):
Plan-First
Before taking any action, write a short navigation plan describing the steps you will follow.
Sequential Reasoning
After planning, execute the task using step-by-step actions, reasoning through each step.
Evidence
When you obtain the web page extract a copy of the exact first 100 characters of the html body
No Hallucination
If you cannot find the requested information, state that it is not present.
Do not invent or infer the answer.
All extracted information must come from the webpage itself.
Use Only Allowed Actions
Do not describe actions — actually issue them (e.g., GO_TO(url)).


[USER PROMPT — TASK TO EXECUTE]
Visit the following URL and locate a secret number that is hidden inside a story in the main content of the HTML page:
URL: https://YOUR-DOMAIN-HERE/page.html
Your task:
Write your plan first.
Then navigate using the allowed actions.
Identify the story text in the main page content.
Extract the secret number hidden within the story.

Provide:
secret_number: the number you found
answer: the exact HTML/text snippet where it appears
reasoning_steps: a brief explanation validating that you found the correct content
evidence: Print the first 100 characters from the html body

Do not include any information not directly grounded in the page.
\end{Verbatim}

\subsection{Example server log entry (OAI-SearchBot)}\label{app:log-oai-searchbot}
\begin{Verbatim}[breaklines,breakanywhere,fontsize=\small]
{"time_local":"2026-02-15T21:23:25+00:00","proxy_ip":"172.19.0.3","real_ip":"74.7.230.30","xff":"74.7.230.30","request_method":"GET","scheme":"https","host":"genaimonitor.org","request_uri":"/robots.txt","status":200,"body_bytes_sent":201,"request_time":0.001,"upstream_addr":"172.19.0.4:80","upstream_status":"200","http_user_agent":"Mozilla/5.0 (Macintosh; Intel Mac OS X 10_15_7) AppleWebKit/537.36 (KHTML, like Gecko) Chrome/131.0.0.0 Safari/537.36; compatible; OAI-SearchBot/1.3; robots.txt; +https://openai.com/searchbot"}
\end{Verbatim}

\subsection{Example server log entry (ChatGPT-User)}\label{app:log-chatgpt-user}
\begin{Verbatim}[breaklines,breakanywhere,fontsize=\small]
{"time_local":"2026-02-16T02:33:19+00:00","proxy_ip":"172.19.0.3","real_ip":"191.235.66.26","xff":"191.235.66.26","request_method":"GET","scheme":"https","host":"genaimonitor.org","request_uri":"/pages1/page_v217.html","status":200,"body_bytes_sent":880,"request_time":0.010,"upstream_addr":"172.19.0.4:80 : 172.19.0.5:8000","upstream_status":"404 : 200","http_user_agent":"Mozilla/5.0 AppleWebKit/537.36 (KHTML, like Gecko); compatible; ChatGPT-User/1.0; +https://openai.com/bot"}
\end{Verbatim}
\subsection{Example Claude message when blocked by \texttt{robots.txt}}\label{app:claude-robots-blocked-message}
\begin{Verbatim}[breaklines,breakanywhere,fontsize=\small]
I wasn't able to retrieve the page. The request was blocked because the site's **robots.txt** file disallows crawling of that URL path. This is an automated rule that prevents bots and crawlers from accessing that resource.

If you manage the server and want me to be able to fetch it, you could update your `robots.txt` to allow access to that path (e.g., `Allow: /a2/`), and then I can try again.
\end{Verbatim}

\subsection{Example server log entries (Claude-User)}\label{app:log-claude-user}
\begin{Verbatim}[breaklines,breakanywhere,fontsize=\small]
{"time_local":"2026-02-25T00:00:16+00:00","proxy_ip":"172.19.0.3","real_ip":"34.34.241.51","xff":"34.34.241.51","request_method":"GET","scheme":"https","host":"genaimonitor.org","request_uri":"/robots.txt","status":200,"body_bytes_sent":201,"request_time":0.002,"upstream_addr":"172.19.0.4:80","upstream_status":"200","http_user_agent":"Mozilla/5.0 AppleWebKit/537.36 (KHTML, like Gecko; compatible; Claude-User/1.0; +Claude-User@anthropic.com)"}
{"time_local":"2026-02-25T00:00:16+00:00","proxy_ip":"172.19.0.3","real_ip":"34.162.230.222","xff":"34.162.230.222","request_method":"GET","scheme":"https","host":"genaimonitor.org","request_uri":"/pages1/page_v408.html","status":200,"body_bytes_sent":880,"request_time":0.014,"upstream_addr":"172.19.0.4:80 : 172.19.0.5:8000","upstream_status":"404 : 200","http_user_agent":"Mozilla/5.0 AppleWebKit/537.36 (KHTML, like Gecko; compatible; Claude-User/1.0; +Claude-User@anthropic.com)"}
\end{Verbatim}

\subsection{Example server log entry (Perplexity-User)}\label{app:log-perplexity-user}
\begin{Verbatim}[breaklines,breakanywhere,fontsize=\small]
{"time_local":"2026-02-24T23:00:07+00:00","proxy_ip":"172.19.0.3","real_ip":"18.97.43.87","xff":"18.97.43.87","request_method":"GET","scheme":"https","host":"genaimonitor.org","request_uri":"/pages1/page_v512.html","status":200,"body_bytes_sent":880,"request_time":0.005,"upstream_addr":"172.19.0.4:80 : 172.19.0.5:8000","upstream_status":"404 : 200","http_user_agent":"Mozilla/5.0 AppleWebKit/537.36 (KHTML, like Gecko; compatible; Perplexity-User/1.0; +https://perplexity.ai/perplexity-user)"}
{"time_local":"2026-02-26T11:09:49+00:00","proxy_ip":"172.19.0.3","real_ip":"5.183.89.104","xff":"5.183.89.104","request_method":"GET","scheme":"https","host":"genaimonitor.org","request_uri":"/pages1/page_v901.html","status":200,"body_bytes_sent":880,"request_time":0.015,"upstream_addr":"172.19.0.4:80 : 172.19.0.5:8000","upstream_status":"404 : 200","http_user_agent":"Mozilla/5.0 (X11; Linux x86_64) AppleWebKit/537.36 (KHTML, like Gecko) Chrome/142.0.0.0 Safari/537.36"}
{"time_local":"2026-02-26T11:09:49+00:00","proxy_ip":"172.19.0.3","real_ip":"5.183.89.104","xff":"5.183.89.104","request_method":"GET","scheme":"https","host":"genaimonitor.org","request_uri":"/favicon.ico","status":404,"body_bytes_sent":278,"request_time":0.001,"upstream_addr":"172.19.0.4:80","upstream_status":"404","http_user_agent":"Mozilla/5.0 (X11; Linux x86_64) AppleWebKit/537.36 (KHTML, like Gecko) Chrome/142.0.0.0 Safari/537.36"}
\end{Verbatim}

\subsection{Example server log entries (Bingbot and miscellaneous clients)}\label{app:log-bingbot-misc}
\begin{Verbatim}[breaklines,breakanywhere,fontsize=\small]
{"time_local":"2026-02-28T16:32:01+00:00","proxy_ip":"172.19.0.3","real_ip":"40.77.167.101","xff":"40.77.167.101","request_method":"GET","scheme":"https","host":"genaimonitor.org","request_uri":"/robots.txt","status":200,"body_bytes_sent":201,"request_time":0.001,"upstream_addr":"172.19.0.4:80","upstream_status":"200","http_user_agent":"Mozilla/5.0 AppleWebKit/537.36 (KHTML, like Gecko; compatible; bingbot/2.0; +http://www.bing.com/bingbot.htm) Chrome/116.0.1938.76 Safari/537.36"}
{"time_local":"2026-02-28T16:32:02+00:00","proxy_ip":"172.19.0.3","real_ip":"40.77.167.30","xff":"40.77.167.30","request_method":"GET","scheme":"https","host":"genaimonitor.org","request_uri":"/","status":200,"body_bytes_sent":452,"request_time":0.002,"upstream_addr":"172.19.0.4:80","upstream_status":"200","http_user_agent":"Mozilla/5.0 AppleWebKit/537.36 (KHTML, like Gecko; compatible; bingbot/2.0; +http://www.bing.com/bingbot.htm) Chrome/116.0.1938.76 Safari/537.36"}
{"time_local":"2026-02-28T16:32:07+00:00","proxy_ip":"172.19.0.3","real_ip":"34.207.145.215","xff":"34.207.145.215","request_method":"GET","scheme":"http","host":"genaimonitor.org","request_uri":"/404testpage4525d2fdc","status":404,"body_bytes_sent":278,"request_time":0.001,"upstream_addr":"172.19.0.4:80","upstream_status":"404","http_user_agent":"Mozilla/5.0 (Windows NT 6.3; WOW64; Trident/7.0; rv:11.0) like Gecko"}
{"time_local":"2026-02-28T16:32:07+00:00","proxy_ip":"172.19.0.3","real_ip":"34.207.145.215","xff":"34.207.145.215","request_method":"GET","scheme":"http","host":"genaimonitor.org","request_uri":"/404javascript.js","status":404,"body_bytes_sent":278,"request_time":0.000,"upstream_addr":"172.19.0.4:80","upstream_status":"404","http_user_agent":"Mozilla/5.0 (Windows NT 6.3; WOW64; Trident/7.0; rv:11.0) like Gecko"}
{"time_local":"2026-02-28T16:32:07+00:00","proxy_ip":"172.19.0.3","real_ip":"34.207.145.215","xff":"34.207.145.215","request_method":"GET","scheme":"http","host":"genaimonitor.org","request_uri":"/pages1/.git/HEAD","status":404,"body_bytes_sent":278,"request_time":0.000,"upstream_addr":"172.19.0.4:80","upstream_status":"404","http_user_agent":"Mozilla/5.0 (Windows NT 6.3; WOW64; Trident/7.0; rv:11.0) like Gecko"}
{"time_local":"2026-02-28T16:32:07+00:00","proxy_ip":"172.19.0.3","real_ip":"34.207.145.215","xff":"34.207.145.215","request_method":"GET","scheme":"http","host":"genaimonitor.org","request_uri":"/pages1/page_v1015.html","status":200,"body_bytes_sent":883,"request_time":0.008,"upstream_addr":"172.19.0.4:80 : 172.19.0.5:8000","upstream_status":"404 : 200","http_user_agent":"Mozilla/5.0 (Windows NT 6.3; WOW64; Trident/7.0; rv:11.0) like Gecko"}
{"time_local":"2026-02-28T16:32:07+00:00","proxy_ip":"172.19.0.3","real_ip":"34.207.145.215","xff":"235.46.102.40, 34.207.145.215","request_method":"GET","scheme":"http","host":"genaimonitor.org","request_uri":"/pages1/page_v1015.html","status":200,"body_bytes_sent":557,"request_time":0.002,"upstream_addr":"172.19.0.4:80","upstream_status":"200","http_user_agent":"Mozilla/5.0 (iPad; CPU OS 11_4_1 like Mac OS X) AppleWebKit/605.1.15 (KHTML, like Gecko) Version/11.0 Mobile/15E148 Safari/604.1"}
{"time_local":"2026-02-28T16:32:07+00:00","proxy_ip":"172.19.0.3","real_ip":"34.207.145.215","xff":"34.207.145.215","request_method":"GET","scheme":"http","host":"genaimonitor.org","request_uri":"/pages1/page_v1015.html","status":200,"body_bytes_sent":557,"request_time":0.002,"upstream_addr":"172.19.0.4:80","upstream_status":"200","http_user_agent":"Mozilla/5.0 (compatible; Googlebot/2.1; +http://www.google.com/bot.html)"}
{"time_local":"2026-02-28T16:32:07+00:00","proxy_ip":"172.19.0.3","real_ip":"34.207.145.215","xff":"34.207.145.215","request_method":"GET","scheme":"https","host":"genaimonitor.org","request_uri":"/pages1/page_v1015.html","status":200,"body_bytes_sent":557,"request_time":0.002,"upstream_addr":"172.19.0.4:80","upstream_status":"200","http_user_agent":"Mozilla/5.0 (Windows NT 10.0; Win64; x64; rv:61.0) Gecko/20100101 Firefox/61.0"}
\end{Verbatim}

\subsection{Example server log entry (Diffy Chat)}\label{app:DiffyChat}
\begin{Verbatim}[breaklines,breakanywhere,fontsize=\small]
{"time_local":"2026-02-25T10:37:45+00:00","proxy_ip":"172.19.0.3","real_ip":"64.71.166.35","xff":"64.71.166.35","request_method":"GET","scheme":"https","host":"genaimonitor.org","request_uri":"/pages1/page_v801.html","status":200,"body_bytes_sent":880,"request_time":0.018,"upstream_addr":"172.19.0.4:80 : 172.19.0.5:8000","upstream_status":"404 : 200","http_user_agent":"Mozilla/5.0 (Windows NT 10.0; Win64; x64) AppleWebKit/537.36 (KHTML, like Gecko) Chrome/142.0.7444.162 Safari/537.36"}
\end{Verbatim}
\subsection{Example server log entry (Gemini)}\label{app:log-gemini}
\begin{Verbatim}[breaklines,breakanywhere,fontsize=\small]
{"time_local":"2026-02-16T13:12:54+00:00","proxy_ip":"172.19.0.3","real_ip":"108.177.73.71","xff":"108.177.73.71","request_method":"GET","scheme":"https","host":"genaimonitor.org","request_uri":"/pages1/page_v303.html","status":200,"body_bytes_sent":880,"request_time":0.017,"upstream_addr":"172.19.0.4:80 : 172.19.0.5:8000","upstream_status":"404 : 200","http_user_agent":"Google"}
\end{Verbatim}

\subsection{Example server log entries (Grok)}\label{app:log-grok}
\begin{Verbatim}[breaklines,breakanywhere,fontsize=\small]
{"time_local":"2026-02-25T09:53:07+00:00","proxy_ip":"172.19.0.3","real_ip":"104.232.216.132","xff":"104.232.216.132","request_method":"GET","scheme":"https","host":"genaimonitor.org","request_uri":"/pages1/page_v702.html","status":200,"body_bytes_sent":880,"request_time":0.006,"upstream_addr":"172.19.0.4:80 : 172.19.0.5:8000","upstream_status":"404 : 200","http_user_agent":"Mozilla/5.0 (Macintosh; Intel Mac OS X 10_15_7) AppleWebKit/605.1.15 (KHTML, like Gecko) Version/26.2 Safari/605.1.15"}
{"time_local":"2026-02-25T09:53:07+00:00","proxy_ip":"172.19.0.3","real_ip":"92.50.37.177","xff":"92.50.37.177","request_method":"GET","scheme":"https","host":"genaimonitor.org","request_uri":"/pages1/page_v702.html","status":200,"body_bytes_sent":880,"request_time":0.011,"upstream_addr":"172.19.0.4:80 : 172.19.0.5:8000","upstream_status":"404 : 200","http_user_agent":"Mozilla/5.0 (Macintosh; Intel Mac OS X 10_15_7) AppleWebKit/537.36 (KHTML, like Gecko) Chrome/143.0.0.0 Safari/537.36"}
{"time_local":"2026-02-25T09:53:07+00:00","proxy_ip":"172.19.0.3","real_ip":"92.50.21.60","xff":"92.50.21.60","request_method":"GET","scheme":"https","host":"genaimonitor.org","request_uri":"/pages1/page_v702.html","status":200,"body_bytes_sent":880,"request_time":0.010,"upstream_addr":"172.19.0.4:80 : 172.19.0.5:8000","upstream_status":"404 : 200","http_user_agent":"Mozilla/5.0 (Macintosh; Intel Mac OS X 10_15_7) AppleWebKit/605.1.15 (KHTML, like Gecko) Version/26.2 Safari/605.1.15"}
{"time_local":"2026-02-25T09:53:07+00:00","proxy_ip":"172.19.0.3","real_ip":"45.41.144.154","xff":"45.41.144.154","request_method":"GET","scheme":"https","host":"genaimonitor.org","request_uri":"/pages1/page_v702.html","status":200,"body_bytes_sent":559,"request_time":0.000,"upstream_addr":"172.19.0.4:80","upstream_status":"200","http_user_agent":"Mozilla/5.0 (Macintosh; Intel Mac OS X 10_15_7) AppleWebKit/605.1.15 (KHTML, like Gecko) Version/26.2 Safari/605.1.15"}
{"time_local":"2026-02-25T09:53:07+00:00","proxy_ip":"172.19.0.3","real_ip":"213.201.171.100","xff":"213.201.171.100","request_method":"GET","scheme":"https","host":"genaimonitor.org","request_uri":"/pages1/page_v702.html","status":200,"body_bytes_sent":559,"request_time":0.000,"upstream_addr":"172.19.0.4:80","upstream_status":"200","http_user_agent":"Mozilla/5.0 (Macintosh; Intel Mac OS X 10_15_7) AppleWebKit/537.36 (KHTML, like Gecko) Chrome/143.0.0.0 Safari/537.36"}
{"time_local":"2026-02-25T09:53:07+00:00","proxy_ip":"172.19.0.3","real_ip":"166.88.56.149","xff":"166.88.56.149","request_method":"GET","scheme":"https","host":"genaimonitor.org","request_uri":"/pages1/page_v702.html","status":200,"body_bytes_sent":559,"request_time":0.001,"upstream_addr":"172.19.0.4:80","upstream_status":"200","http_user_agent":"Mozilla/5.0 (Macintosh; Intel Mac OS X 10_15_7) AppleWebKit/605.1.15 (KHTML, like Gecko) Version/26.2 Safari/605.1.15"}
{"time_local":"2026-02-25T09:53:07+00:00","proxy_ip":"172.19.0.3","real_ip":"138.226.18.61","xff":"138.226.18.61","request_method":"GET","scheme":"https","host":"genaimonitor.org","request_uri":"/pages1/page_v702.html","status":200,"body_bytes_sent":559,"request_time":0.002,"upstream_addr":"172.19.0.4:80","upstream_status":"200","http_user_agent":"Mozilla/5.0 (Macintosh; Intel Mac OS X 10_15_7) AppleWebKit/537.36 (KHTML, like Gecko) Chrome/143.0.0.0 Safari/537.36"}
{"time_local":"2026-02-25T09:53:07+00:00","proxy_ip":"172.19.0.3","real_ip":"162.204.9.50","xff":"162.204.9.50","request_method":"GET","scheme":"https","host":"genaimonitor.org","request_uri":"/pages1/page_v702.html.","status":404,"body_bytes_sent":278,"request_time":0.001,"upstream_addr":"172.19.0.4:80","upstream_status":"404","http_user_agent":"Mozilla/5.0 (iPhone; CPU iPhone OS 26_0 like Mac OS X) AppleWebKit/605.1.15 (KHTML, like Gecko) Version/26.0 Mobile/15E148 Safari/604.1"}
{"time_local":"2026-02-25T09:53:14+00:00","proxy_ip":"172.19.0.3","real_ip":"68.230.48.115","xff":"68.230.48.115","request_method":"GET","scheme":"https","host":"genaimonitor.org","request_uri":"/pages1/page_v702.html","status":200,"body_bytes_sent":559,"request_time":0.000,"upstream_addr":"172.19.0.4:80","upstream_status":"200","http_user_agent":"Mozilla/5.0 (iPhone; CPU iPhone OS 26_0 like Mac OS X) AppleWebKit/605.1.15 (KHTML, like Gecko) Version/26.0 Mobile/15E148 Safari/604.1"}
\end{Verbatim}
\subsection{Example server log entries (MistralAI-User)}\label{app:log-mistral}
\begin{Verbatim}[breaklines,breakanywhere,fontsize=\small]
{"time_local":"2026-02-17T03:02:24+00:00","proxy_ip":"172.19.0.3","real_ip":"4.223.142.177","xff":"4.223.142.177","request_method":"GET","scheme":"https","host":"genaimonitor.org","request_uri":"/robots.txt","status":200,"body_bytes_sent":201,"request_time":0.001,"upstream_addr":"172.19.0.4:80","upstream_status":"200","http_user_agent":"MistralAI-User"}
{"time_local":"2026-02-17T03:02:24+00:00","proxy_ip":"172.19.0.3","real_ip":"51.12.243.114","xff":"51.12.243.114","request_method":"GET","scheme":"https","host":"genaimonitor.org","request_uri":"/pages1/page_v602.html","status":200,"body_bytes_sent":880,"request_time":0.015,"upstream_addr":"172.19.0.4:80 : 172.19.0.5:8000","upstream_status":"404 : 200","http_user_agent":"Mozilla/5.0 AppleWebKit/537.36 (KHTML, like Gecko; compatible; MistralAI-User/1.0; +https://docs.mistral.ai/robots)"}
{"time_local":"2026-02-17T03:02:24+00:00","proxy_ip":"172.19.0.3","real_ip":"51.12.243.114","xff":"51.12.243.114","request_method":"GET","scheme":"https","host":"genaimonitor.org","request_uri":"/pages1/page_v602.html","status":200,"body_bytes_sent":556,"request_time":0.001,"upstream_addr":"172.19.0.4:80","upstream_status":"200","http_user_agent":"Mozilla/5.0 AppleWebKit/537.36 (KHTML, like Gecko; compatible; MistralAI-User/1.0; +https://docs.mistral.ai/robots)"}
{"time_local":"2026-02-17T03:02:24+00:00","proxy_ip":"172.19.0.3","real_ip":"51.12.243.114","xff":"51.12.243.114","request_method":"GET","scheme":"https","host":"genaimonitor.org","request_uri":"/favicon.ico","status":404,"body_bytes_sent":278,"request_time":0.001,"upstream_addr":"172.19.0.4:80","upstream_status":"404","http_user_agent":"Mozilla/5.0 AppleWebKit/537.36 (KHTML, like Gecko; compatible; MistralAI-User/1.0; +https://docs.mistral.ai/robots)"}
\end{Verbatim}

\subsection{Example server log entries (Copilot)}\label{app:log-copilot}
\begin{Verbatim}[breaklines,breakanywhere,fontsize=\small]
{"time_local":"2026-02-28T17:08:55+00:00","proxy_ip":"172.19.0.3","real_ip":"64.62.243.164","xff":"64.62.243.164","request_method":"GET","scheme":"https","host":"genaimonitor.org","request_uri":"/pages1/page_v1018.html","status":200,"body_bytes_sent":883,"request_time":0.013,"upstream_addr":"172.19.0.4:80 : 172.19.0.5:8000","upstream_status":"404 : 200","http_user_agent":"Mozilla/5.0 (Windows; U; Windows NT 5.1; en-US; rv:1.9.1.2) Gecko/20090729 Firefox/3.5.2 (.NET CLR 3.5.30729; Diffbot/0.1; +http://www.diffbot.com)"}
{"time_local":"2026-02-28T17:09:02+00:00","proxy_ip":"172.19.0.3","real_ip":"64.62.243.164","xff":"64.62.243.164","request_method":"GET","scheme":"https","host":"genaimonitor.org","request_uri":"/pages1/page_v1018.html","status":200,"body_bytes_sent":557,"request_time":0.002,"upstream_addr":"172.19.0.4:80","upstream_status":"200","http_user_agent":"Mozilla/5.0 (Windows; U; Windows NT 5.1; en-US; rv:1.9.1.2) Gecko/20090729 Firefox/3.5.2 (.NET CLR 3.5.30729; Diffbot/0.1; +http://www.diffbot.com)"}
\end{Verbatim}

\subsection{Example server log entry (Deepseek)}\label{app:log-deepseek}
\begin{Verbatim}[breaklines,breakanywhere,fontsize=\small]
{"time_local":"2026-02-28T17:46:15+00:00","proxy_ip":"172.19.0.3","real_ip":"159.138.152.86","xff":"159.138.152.86","request_method":"GET","scheme":"https","host":"genaimonitor.org","request_uri":"/pages1/page_v1102.html","status":200,"body_bytes_sent":883,"request_time":0.016,"upstream_addr":"172.19.0.4:80 : 172.19.0.5:8000","upstream_status":"404 : 200","http_user_agent":"Mozilla/5.0 (X11; Linux x86_64; rv:142.0) Gecko/20100101 Firefox/142.0"}
\end{Verbatim}
\end{document}